\documentclass[a4paper,twocolumn,superscriptaddress,9pt]{revtex4-1}

\usepackage{xcolor}
\usepackage{hyperref}
\hypersetup{
	colorlinks,
	linkcolor={red!90!black},
	citecolor={black!10!blue},
	urlcolor={blue!80!black}
}

\usepackage{tikz}
\usepackage{amssymb}
\usepackage{amsmath}
\usepackage{graphicx}
\usepackage{multirow}
\usepackage[utf8]{inputenc}
\usepackage{color}

\renewcommand{\d}{{d}}

\renewcommand{\r}{\textcolor{red}}
\newcommand{\p}{\partial}
\newcommand{\f}{\frac}

\newcommand{\br}{\mathbf{r}}

\renewcommand{\u}{\mathbf{u}^{\text{ext}}}

\newcommand{\bk}{\mathbf{k}}
\newcommand{\bq}{\mathbf{q}}

\renewcommand{\theequation}{\arabic{equation}}
\renewcommand{\thefigure}{\arabic{figure}}
\renewcommand{\thesection}{\Roman{section}}

\renewcommand{\d}{\mathrm{d}}

\renewcommand{\r}{\textcolor{red}}

\renewcommand{\u}{\mathbf{u}^{\text{ext}}}

\renewcommand{\a}{\alpha}
\renewcommand{\b}{\beta}
\renewcommand{\d}{\mathrm{d}}
\renewcommand{\t}{\tau}
\newcommand{\D}{\mathcal{D}}
\renewcommand{\S}{\Sigma}
\renewcommand{\l}{\lambda}

\newcommand{\DD}{\Delta}

\newcommand{\bv}{\mathbf{v}}
\newcommand{\nf}{\mathbf{f}}
\newcommand{\mF}{\mathcal{F}}

\newcommand{\lammct}{\lambda_\text{MCT}}
\newcommand{\lamc}{\lambda_\text{C}}

\begin{document}

\title{Mode-coupling theory for aging in active glasses: relaxation dynamics and evolution towards steady state}
	\author{Soumitra Kolya}
	\email{soumitrakolya@tifrh.res.in}
	\affiliation{Tata Institute of Fundamental Research, Gopanpally Village, Hyderabad - 500046, India}
	\author{Nir S. Gov}
	\email{nir.gov@weizmann.ac.il}
	\affiliation{Department of Chemical and Biological Physics, Weizmann Institute of Science, Rehovot 7610001, Israel}
    \author{Saroj Kumar Nandi}
    \email{saroj@tifrh.res.in}
    \affiliation{Tata Institute of Fundamental Research, Gopanpally Village, Hyderabad - 500046, India}
	   
\begin{abstract} 
Aging refers to the evolution of system properties with waiting time $t_w$. It is a key feature of glassy dynamics. Recent experiments have demonstrated aging in biological systems that are inherently active with a magnitude of self-propulsion force $f_0$ and a persistence time $\tau_p$. Thus, what governs the aging dynamics in these active systems has fundamental importance. We formulate a generic mode-coupling theory (MCT) of active glasses to address this question. The aging solutions of the theory show that the two-point correlation function decays more slowly with growing $t_w$, and the relaxation time $t_r$ increases. The activity-modification of the MCT critical point, $\lamc$, has profound significance for active aging: the quench distance from $\lamc$ governs aging and determines $\delta$, where $t_r\sim t_w^\delta$. $\delta$ decreases with increasing $f_0$, in agreement with existing simulations. However, the variation with $\tau_p$ depends on the nature of activity. Our work has fundamental theoretical implications for active glasses and paves the way for a deeper understanding of the aging dynamics in biological systems.

\end{abstract}
\maketitle	

\section{Introduction}

Aging is a fundamental characteristic of glassy dynamics 
\cite{cugliandolo1993analytical,nandi2012glassy,nandi2016glass,lunkenheimer2005glassy,abou2001aging}; it ``{\it refers to structural relaxation of the glassy state toward the metastable equilibrium amorphous state}'' \cite{hodge1995} and affects nearly all physical properties of the system. Many recent experiments~\cite{fabry2001scaling,zhou2009universal,garcia2015physics,angelini2011glass} have revealed glassy dynamics in various biological systems, exhibiting key characteristics of glasses, such as complex stretched-exponential relaxation~\cite{park2015unjamming,malinverno2017endocytic,atia2018geometric}, a non-Gaussian displacement distribution~\cite{bursac2005cytoskeletal,giavazzi2018}, dynamic heterogeneity~\cite{angelini2011glass,cerbino2021,malinverno2017endocytic,park2015unjamming}, sharply growing relaxation time and viscosity \cite{nishizawa2017,park2015unjamming}, etc. 
Examples include intracellular phase-separated biomolecular condensates~\cite{jawerth2020protein,alshareedah2021programmable,takaki2023theory}, the cellular cytoplasm~\cite{fabry2001scaling,bursac2005cytoskeletal,deng2006fast}, confluent epithelial monolayers and tissues~\cite{angelini2011glass,park2015unjamming,atia2018geometric,sadhukhan2021theory}, and collections of organisms~\cite{takatori2020motility,lama2024emergence} and synthetic systems \cite{klongvessa2019,klongvessa2022,arora2022,vijay2024}. In addition, recent experiments have shown that several of these systems also show the aging dynamics \cite{atia2018geometric,park2015unjamming,bursac2005cytoskeletal,jawerth2020protein,alshareedah2021programmable,takaki2023theory}, here the system properties change with the waiting time, $t_w$. In the context of the biological systems, we define the waiting time $t_w$ since the system was prepared or perturbed.
The glassy dynamics in these systems are crucial for various biological processes, such as wound healing \cite{malinverno2017endocytic,poujade2007collective,brugues2014forces}, embryogenesis \cite{friedl2009collective}, and cancer progression \cite{tambe2011collective,malmi2018cell}; therefore, it is imperative to understand the changing system properties with the waiting time.

Our focus here is the physical aging, the slow evolution of structure and mechanical properties and not the chemical reactions rate, of the biological systems. However, biological systems are immensely complex, with diverse and novel control parameters and phenomenologies, making it challenging to develop a comprehensive theoretical framework. Thus, simplified model systems, containing only specific details of the biological systems, have nevertheless been instrumental in understanding the roles of different aspects of these systems \cite{sadhukhan2024perspective}. One crucial feature of the biological systems compared to inert, passive systems is their activity, where the constituents have a self-propulsion force of magnitude $f_0$ and persistence time $\tau_p$, and they can also be strongly related to each other \cite{maiuri2015actin,wortel2021}. Dense systems of self-propelled particles exhibit glassy dynamics, and they are known as active glasses. These models have provided crucial insights into the role of activity in the glassy dynamics 
\cite{sadhukhan2024perspective,janssen2019active,berthier2019glassy}. Analytical theories \cite{berthier2013,szamel2015,liluashvili2017,feng2017,debets2022,nandi2017nonequilibrium,nandi2018random,paul2023dynamical,kolya2024active} and simulation studies \cite{berthier2014,flenner2016,mandal2016active,debets2021,mandal2021study,keta2023intermittent,keta2022disordered} of the simplified models show that activity has nontrivial effects on the glassy dynamics; for example, it can modify the glass transition point, lead to reentrant dynamics, and modulate fragility \cite{sadhukhan2024perspective,pareek2025,berthier2013,nandi2017nonequilibrium,flenner2016,mandal2016active,debets2021}.

Despite these theoretical advances, how activity influences the non-stationary state, i.e., aging dynamics, remains unknown. To the best of our knowledge, only two simulation studies to date \cite{mandal2020multiple,janzen2022aging} have focused on the aging dynamics in dense systems of active Brownian particles (ABPs). For an athermal active system, Ref.~\cite{mandal2020multiple} has shown that aging in the presence of activity resembles that of a passive thermal system when $\tau_p\to0$; nontrivial behavior emerges only at large $\tau_p$. On the other hand, Ref.~\cite{janzen2022aging} studied a thermal ABP model and found activity-dependent aging even when $\tau_p$ is small. Aging in passive glasses has been extensively studied experimentally~\cite{bonn2002laponite,di2011signatures,lunkenheimer2005glassy,ramos2001ultraslow,abou2001aging,riechers2022predicting}, through simulations~\cite{kob1997aging,simha1984molecular}, and within the theoretical frameworks of mode-coupling theory (MCT)~\cite{cugliandolo1993analytical,nandi2012glassy,nandi2016glass} and Random First Order Transition theory~\cite{peter2009spatiotemporal,lubchenko2004theory}. However, these theoretical frameworks have not been extended to the aging dynamics in active glasses.

In this work, we study the aging dynamics in active glasses of self-propelled particles within the framework of mode-coupling theory (MCT). Note that there are many different routes to obtain the steady-state forms of active MCT \cite{szamel2015,nandi2017nonequilibrium,liluashvili2017,debets2022,feng2017}. However, the aging dynamics for passive glasses to date has been captured within MCT only via the field-theoretic approach \cite{nandi2012glassy,nandi2016glass,cugliandolo1993analytical}. Therefore, here, we focus on this specific approach to derive the nonequilibrium theory. We have developed a suitable numerical algorithm to solve our non-stationary active MCT equations. We provide the details of the numerical algorithm in the supplementary material (SM). We demonstrate that both the distance of the quench from the critical point and the activity-modification of the critical points play crucial roles in the aging dynamics in the presence of activity. We organize the rest of the paper as follows: We provide a brief overview of the derivation of the active aging MCT, in Sec. \ref{derivation}. We present the behavior of the two-point correlation function with $t_w$ and the nature of the aging dynamics in Sec. \ref{result_aging}, and then show in Sec. \ref{dist_lamc} that the distance of the quench from the modified critical point, $\lamc$, governs the aging dynamics. We demonstrate in Sec. \ref{evol_SS} that the stationary state of the aging MCT, when the quench is in the liquid state, agrees with the steady state active MCT. We spell out the predictions of the theory and compare them with existing simulations in Sec. \ref{comp_pred}. We conclude the paper in Sec. \ref{disc} with a discussion of our results and how they relate to the aging dynamics in biological systems.

\section{Non-stationary MCT for an aging active system}
\label{derivation}
We start with fluctuating hydrodynamic equations for an active system. The continuity equations for the particle density, $\rho(\br,t)$, and the momentum density, $\rho(\br,t)\bv(\br,t)$, where $\bv(\br,t)$ is the velocity field at position $\br$ and time $t$, are 

\begin{align}
	&\f{\p \rho(\br,t)}{\p t}=-\nabla\cdot[\rho(\br,t)\bv(\br,t)] \label{cont_rho}\\
	&\f{\p(\rho\bv)}{\p t}+\nabla\cdot(\rho\bv \bv)=\eta\nabla^2\bv+(\zeta+\eta/3)\nabla\nabla\cdot\bv \nonumber\\ &\hspace{3cm}-\rho\nabla\f{\delta \mF}{\delta \rho}+\nf_T+\nf_A,\label{cont_momentum}
\end{align}
where $\zeta$ and $\eta$ are the bulk and shear viscosities, respectively, while $\nf_T$ and $\nf_A$ denote the thermal and active noises, respectively. $\nf_T$ has zero mean and variance
\begin{equation}
	\langle \nf_T({\bf 0},t)\nf_T'({\bf r},t)\rangle=-2k_BT[\eta {\bf I}\nabla^2+(\zeta+\f{\eta}{3})\nabla\nabla]\delta({\bf r})\delta(t), \nonumber
\end{equation}
where $\nf_T'$ denotes the transpose, ${\bf I}$ is the unit tensor, and $k_B T$, the Boltzmann constant times the temperature. Activity, in the form of self-propulsion, enters the theory via the active noise  $\nf_A$, it also has zero mean, but variance
\begin{equation}\label{activenoise}
	\langle \nf_A({\bf 0},t)\nf_A'({\bf r},t)\rangle=2\DD(t)\delta(\br),
\end{equation}
where $\DD(t)$ depends on the type of activity. Consistent with the forms of activity that have been used in the simulation studies of active glasses \cite{berthier2014,flenner2016,mandal2016active,debets2022}, we have ignored the spatial correlation. For the active Brownian particles (ABP), $\DD(t) = f_0^2 \exp(-t/\tau_p)$,  while for active Ornstein-Uhlenbeck particles (AOUP), $\DD(\textcolor{black}{t})=(f_0^2/\tau_p)\exp(-t/\tau_p)$. Here, $f_0$ is the self-propulsion force and $\tau_p$ is the persistence time. Note that the active noise does not satisfy any fluctuation-dissipation relation as it drives the system out of equilibrium. We treat the system in the limit of small activity where the deviation from the equilibrium is not large \cite{fodor2016,nandi2017nonequilibrium}. In this limit, activity works as a perturbation to the passive system.
$\mathcal{F}$ represents the free-energy functional of the passive system, we chose the Ramakrishnan-Yussouff functional \cite{ramakrishnan1979first}:
\begin{align}
	\beta \mF[\rho]&=\int_\br \rho(\br,t)\left[\ln\f{\rho(\br,t)}{\rho_0}-1\right]\nonumber\\
	&-\f{1}{2}\int_{\br,\br'}\delta\rho(\br,t)c(\br-\br')\delta\rho(\br',t),
\end{align}
where, $\beta = 1/k_B T$ and $\rho_0 = \rho(\br,t) - \delta\rho(\br,t)$ is the average density, with $\delta\rho(\br,t)$ being the density fluctuation, $c(\br - \br')$ is the direct correlation function, and $\int_\br \equiv \int \d\br$.

As we are interested in the glassy dynamics, we want to write the equations in terms of the slow variables, $\rho(\br,t)$. Therefore, we linearlize the fast variable $\bv(\br,t)$ in Eqs. (\ref{cont_rho}) and (\ref{cont_momentum}) by neglecting higher order terms in $\bv(\br,t)$. We then take divergence of Eq. (\ref{cont_momentum}) and substitute $\nabla \cdot \bv$ using the linearized form of Eq. (\ref{cont_rho}). 
 Next, we take a Fourier transform and obtain the equation of motion for $\delta\rho_\bk(t)$, at wave vector $\bk$, as
\begin{align}\label{kdepeq}
	D_Lk^2\f{\p\delta\rho_\bk(t)}{\p t}&+\f{k^2k_BT}{S_k}\delta\rho_\bk(t)= ik \hat{f}_T^L(t)+ik \hat{f}_A^L(t)\nonumber\\
	&+\f{k_BT}{2}\int_\bq \mathcal{V}_{k,q} \delta\rho_\bq(t) \delta\rho_{\bk-\bq}(t),
\end{align}
where $\mathcal{V}{k,q} = \bk \cdot [\bq c_q + (\bk - \bq) c_{k-q}]$, $\hat{f}_T^L$ and $\hat{f}_A^L$ denote the longitudinal components of the Fourier transforms of $\nf_T$ and $\nf_A$, respectively, and $D_L = (\zeta + 4\eta/3)/\rho_0$. $S_k = 1/(1 - \rho_0 c_k)$ is the static structure factor. The above equation gives the starting point for a field-theoretic derivation of MCT \cite{castellani2005spin,reichman2005mode,nandi2012glassy,nandi2016glass}.

Activity provides a separation of time-scale with the thermal noise, therefore, we define the correlation function, $C_k(t,t_w) = \langle\delta\rho_k(t) \delta\rho_{-k}(t_w)\rangle$, and the response function, $R_k(t,t_w) = \langle \partial \delta\rho_k(t) / \partial \hat{f}_T^L(t_w) \rangle$, and from Eq. (\ref{kdepeq}), obtain their equations of motion \cite{nandi2017nonequilibrium} as:
\begin{align}\label{kdepeq1}
	\f{\p C_k(t,t_w)}{\p t} &=-\mu_k(t)C_k(t,t_w)+\int_0^{t_w}\d s\D_k(t,s)R(t_w,s) \nonumber\\
	&+\int_0^t\d s \S_k(t,s)C_k(s,t_w)+2TR_k(t_w,t)\\
	\f{\p R_k(t,t_w)}{\p t} &=-\mu_k(t)R_k(t,t_w) \nonumber\\
	&+\int_{t_w}^t \d s\S_k(t,s)R_k(s,t_w)+\delta(t-t_w) \label{kdepeq2}\\
	\mu_k(t) = T&R_k(0)+\int_0^t \d s[\D_k(t,s)R_k(t,s)+\S_k(t,s)C_k(t,s)], \nonumber
\end{align}
where $\S(t,s)= \kappa_1^2 \int_{\bf q} \mathcal{V}_{k,q}^2 C_{k-q}(t,s)R_q(t,s)$ and $	\D_k(t,s)$ $=\f{\kappa_1^2}{2} \int_{\bf q} \mathcal{V}_{k,q}^2 C_q(t,s)C_{k-q}(t,s)+\kappa_2^2\Delta_k(t-s)$, $\kappa_1 = k_B T / (D_L k^2)$ and $\kappa_2 = 1/D_L$. Equations (\ref{kdepeq1}–\ref{kdepeq2}) represent the nonequilibrium, non-stationary MCT for an active system.

Solving the wave vector dependent equations numerically is impractical even for the steady state MCT \cite{nandi2017nonequilibrium}. Therefore, we schematicize them, write the equations for a specific $k=k_\text{max}$, where the structure factor has the first maximum, $S_{k_\text{max}}$, and then throw away the wave vector dependence. Thus, we obtain the equations of motion for $C(t,t_w) \equiv C_{k=k_{\text{max}}}(t,t_w)/S_{k_\text{max}}$ and $R(t,t_w) \equiv R_{k=k_{\text{max}}}(t,t_w)/S_{k_\text{max}}$ as
\begin{align}
	\f{\p C(t,t_w)}{\p t} =&-\mu(t)C(t,t_w)+\int_0^{t_w}\d s\D(t,s)R(t_w,s) \nonumber\\
	+\int_0^t\d s &\S(t,s)C(s,t_w)+2TR(t_w,t)\label{correq1} \\
	\f{\p R(t,t_w)}{\p t} =&-\mu(t)R(t,t_w) +\delta(t-t_w)  +\int_{t_w}^t \d s\S(t,s)R(s,t_w) \label{responseeq1}\\
	\mu(t) = T&+\int_0^t \d s[\D(t,s)R(t,s)+\S(t,s)C(t,s)] , \nonumber
\end{align}
where $\D_{k = k_{max}}(t,s)\equiv \D(t,s)=2\l C^2(t,s)+\DD(t-s)$ and $\S_{k = k_{max}}(t,s)\equiv \Sigma(t,s)=4\l C(t,s)R(t,s)$. We have the control parameter $\lambda$ in the schematic theory above as $\lambda=(\kappa_1/2S_{k_\text{max}})[\int_\mathbf{q} \mathcal{V}_{k,q}^2S_qS_{k-q}]_{k=k_\text{max}}$.
$\lambda$ contains the information of various control parameters, such as density or $T$, via $\kappa_1$ and the static properties of the system in the form of the static structure factor and the direct correlation function.

Within our theory, activity enters via $\Delta(t)$ whose form will depend on the specific type of activity (ABP vs AOUP). The schematic form provides meaningful insights into the glassy dynamics as it gives the correct time evolution, the primary focus in glassy systems. For the purpose of numerical advantage, we compute the integrated response function defined as $F(t, t_w) = - \int_{t_w}^{t} R(t, s)\, ds$, instead of response function (see SM for the evolution equation of $F(t,t_w)$). The computation of $F(t,t_w)$ is numerically preferred as it has less fluctuations compared to $R(t, t_w)$. We numerically solve the dynamical equations for $C(t, t_w)$ and $F(t, t_w)$ with initial conditions $C(t = t_w, t_w) = 1$ and $F(t = t_w, t_w) = 0$. Equations (\ref{correq1}) and (\ref{responseeq1}) describe the evolution of a system from a very high $T$ (or small $\lambda$) liquid phase after a sudden quench to a particular value of $\lambda$ in the presence of activity. However, even at this level of simplification, the schematic MCT for active aging is challenging for a numerical solution, and the existing algorithms will not work. We have now developed the algorithm for solving these equations (see SM for details) and present the results below.

\begin{figure}
	\includegraphics[height=9.4cm,width= 8cm]{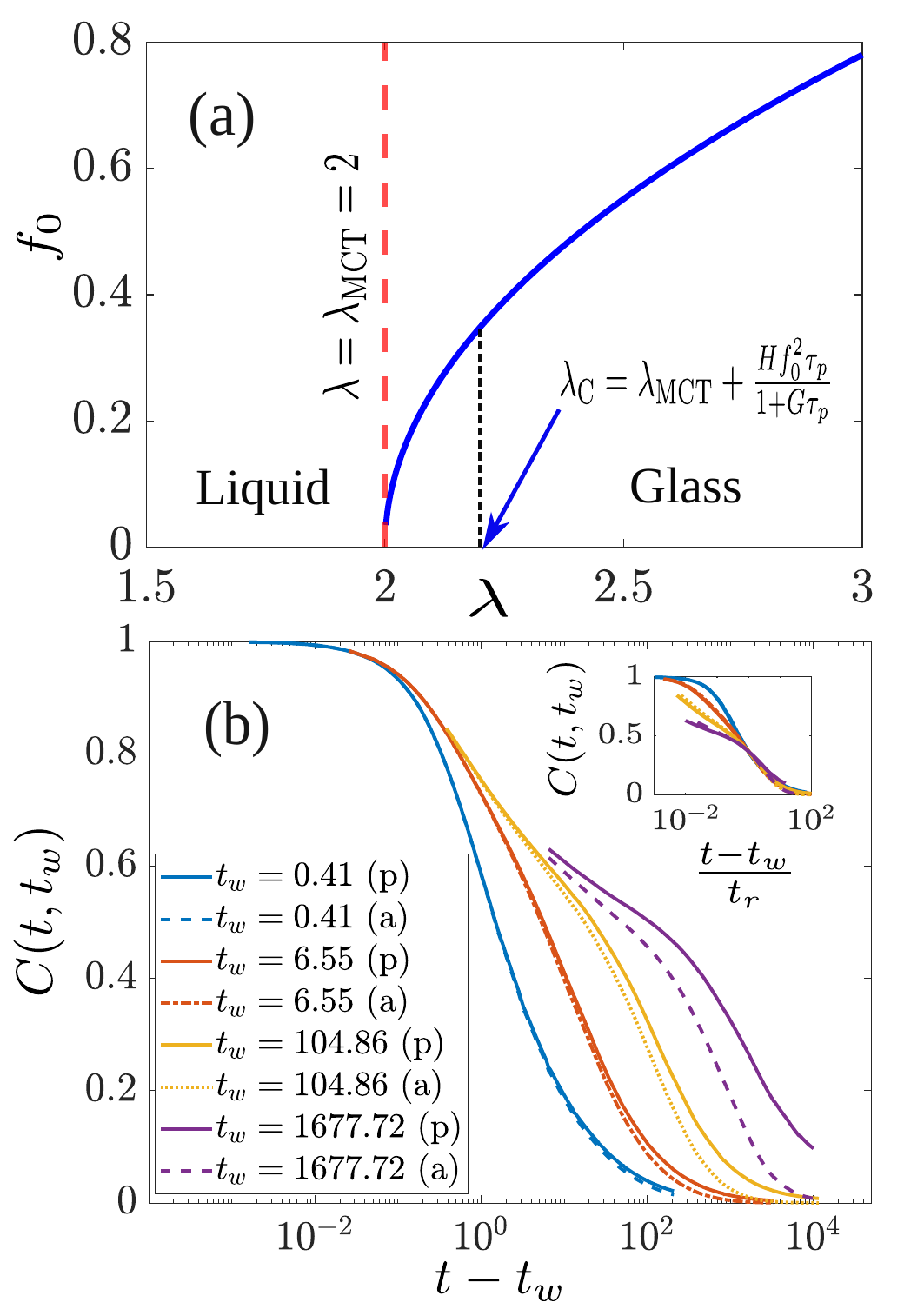}
	\caption{ (a)  A liquid-glass phase diagram of the active system at a fixed $\tau_p$. $\lammct=2$ for the passive system and it increases as $\lamc=\lammct+Hf_0^2\tau_p/(1+G\tau_p)$ where $H$ and $G$ are constants. For the active system, the aging continues forever when the quench is above $\lamc$. The distance from $\lamc$ governs the aging dynamics. As activity modifies $\lamc$, active systems age faster. \textcolor{black}{For $f_0 = 0$, the critical point $\lambda = \lammct = 2$ separate the glass from liquid, for a non zero $f_0$ the critical point shifted to $\lambda_{\text{C}}$ as shown using blue arrow. The small dotted line represents the postion of $\lamc$ for a perticular value of $f_0$.} (b) $C(t,t_w)$ as a function of $(t- t_w)$ for both a passive thermal system (solid lines) and an active thermal system (dotted lines) for a quench to $\lambda=2.01$. The self-propulsion parameters are $f_0=0.2$ and $\tau_p = 2$. {\bf Inset}: Rescaling the time difference $(t - t_w)$ by the relaxation time $t_r$ leads to a data collapse in the $\alpha$-regime.}
	
	\label{fig1}
\end{figure}

\section{results}

\subsection{Aging dynamics}
\label{result_aging}
The aging dynamics refers to the $t_w$ dependence of $C(t,t_w)$ and $F(t,t_w)$. To show the predictions of the theory, we must solve Eqs. (\ref{correq1}) and (\ref{responseeq1}) for all times on a two-dimensional time grid. Moreover, the decay of $C(t,t_w)$ is faster at short times and slower at longer times. This characteristic makes the numerical solution extremely challenging even for the passive system. Additional complications arise for active aging as the evaluation of the activity terms requires $C(t,t_w)$ and $F(t,t_w)$ at all $t$ and $t_w$. We have now developed an algorithm that allows the numerical solution of the aging dynamics in active systems. We present the details of the algorithm in the SM. For clarity of the presentation, we mostly focus on the active aging of the ABP system in this work; we present only some results for the AOUP system in Fig. \ref{predictions}(c) below.
\begin{figure}
	\includegraphics[width=8cm]{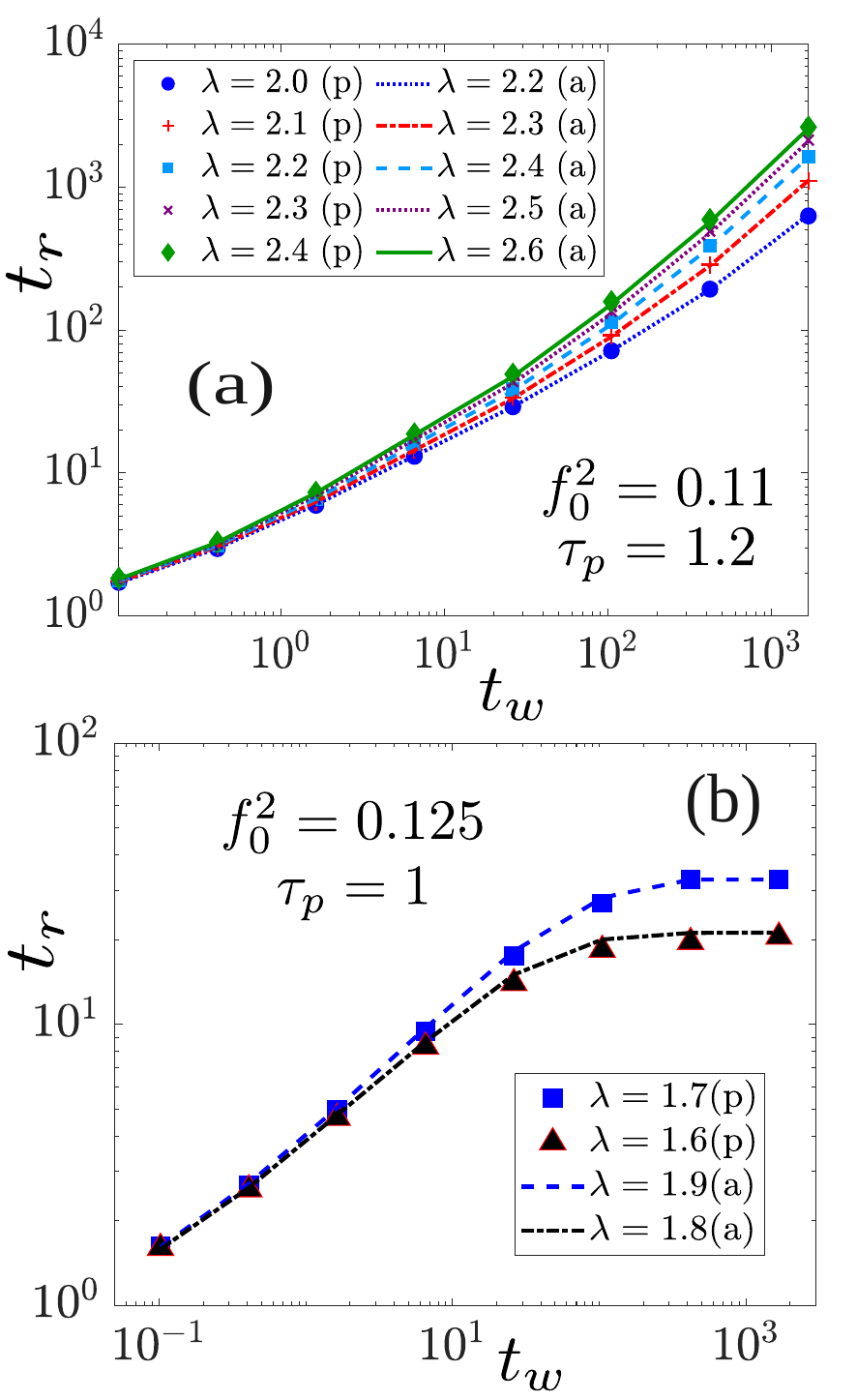}
	\caption{Distance from $\lamc$ governs aging dynamics. \textcolor{black}{The condition for these plots is $\delta \lambda_c = 0.2$}
		(a) Relaxation	time, $t_r(t_w)$, as a function of $t_w$ for the passive and active systems with various quench values of $\lambda$, as shown in the figure. We have $f_0^2=0.11$ and $\tau_p=1.2$ for the active system, giving $\lamc=2.2$. The symbols give $t_r$ for the passive system and lines for the active system. \textcolor{black}{Within the first bracket, `p' denotes the quench value $\lambda$ for the passive system, while `a' denotes the corresponding value for the active system.}
		(b) $t_r$ as a function of $t_w$ saturates to a finite value when the quench is below the transition point. $t_r(t_w)$ for the passive system agrees with that of the active system when the distances from the corresponding critical points are the same.}
	\label{aging}
\end{figure}

For concreteness of the aging protocol, we keep the activity parameters fixed and perform a quench in $\lambda$, which is equivalent to a quench in $T$ or density. Janzen and Janssen have shown in their simulations that a quench in $T$ at fixed activity parameters and a quench in activity at fixed $T$ are equivalent when the final parameters are the same \cite{janzen2022aging}. Note that the non-stationary MCT, Eqs. (\ref{correq1}) and (\ref{responseeq1}), describes the evolution of a system starting from an infinite $T$ or $\lambda\to0$ initial condition towards the glassy state at a specific $\lambda$ after an infinitely rapid quench \cite{nandi2012glassy,nandi2016glass,cugliandolo1993analytical,kimlatz}.
We set $T=1$ and present the results in terms of the final quench value of $\lambda$. Several past works have shown that activity modifies the MCT transition point \cite{berthier2013,berthier2014,mandal2016active,flenner2016,nandi2017nonequilibrium,nandi2018random}. We denote the transition point for the passive system as $\lammct$ and that for the active system as $\lamc$. Considering a single active trapped particle in a confining medium, Ref. \cite{nandi2017nonequilibrium} provided an analytical form for this modified critical point that agrees well with the numerical solution of the steady-state active MCT. For the ABP system, we have \cite{nandi2017nonequilibrium,nandi2018random},
\begin{equation}
	\lambda_{\text{C}} = \lammct + \frac{Hf_0^2 \tau_p}{1 + G \tau_p},
		\label{eq10}
\end{equation}
where $G$ and $H$ are constants. Here we define $\delta \lambda_c = Hf_0^2\tau_p/(1+G\tau_p)$. We show below that this modified critical point has more profound significance for the aging dynamics. From our numerical results, we find $H=3.35$ and $G=1.05$ \textcolor{black}{within the schematic MCT of ABP model}. Figure \ref{fig1}(a) shows the critical line at a fixed $\tau_p$ in the $(f_0-\lambda)$ plane. The dotted vertical line shows $\lamc$ for a specific $(f_0,\tau_p)$. If the quench is above $\lamc$, the aging continues forever. By contrast, if the quench is below $\lamc$, the system reaches steady-state after the initial aging.

We show the evolution of $C(t,t_w)$ as a function of $(t-t_w)$, obtained from the numerical solutions of the aging MCT, Eqs. (\ref{correq1}) and (\ref{responseeq1}), in Fig. \ref{fig1}(b) for an active (dashed lines) and a passive (solid lines) system for a quench to $\lambda=2.01$. 
\textcolor{black}{We have $f_0 = 0.2$ and $\tau_p = 2$ for the active system, for which the corresponding $\lambda$ lies below, but close to, $\lambda_c$.} Unlike in the steady state, $C(t,t_w)$ is no longer a function of $(t-t_w)$ alone, it explicitly depends on $t_w$, signifying aging. We find that the decay of $C(t,t_w)$ becomes faster for the active system, indicating that activity makes the aging faster; this is consistent with the simulations of Ref. \cite{janzen2022aging}. We define the relaxation time $t_r$ via $C(t, t_w) = 1/e$. The inset of Fig.~\ref{fig1}(b) shows data collapse to a master curve for both the passive and active systems in the long-time $\alpha$-regime when we rescale time with $t_r$. This data collapse shows that, similar to the passive system, the active system also exhibits simple aging.

\subsection{Distance from $\lamc$ governs aging }
\label{dist_lamc}
We now show that the aging dynamics in active glasses has similarities with that in passive systems and demonstrate that the former is governed by the distance of the quench from the modified critical point, $\lamc$. We first focus on the regime where the quench is above the MCT critical points of the corresponding system. We quench the passive system above $\lammct$ to different values of $\lambda$. \textcolor{black}{Figure \ref{aging}(a) shows $t_r$ as a function of $t_w$ for various $\lambda$. For the active system, we quench it to a $\lambda$ above $\lamc$. In Figure\ref{aging}(a), we use $f_0^2=0.11$ and $\tau_p=1.2$ to show $t_r$ as a function of $t_w$ for the active system. Note that for these values of $f_0$ and $\tau_p$, we have from equation (\ref{eq10}) $\delta \lambda_c \simeq 0.2$.} Figure \ref{aging}(a) shows that the curve with a specific value of $(\lambda-\lammct)$ for the passive system (symbols) overlaps with that of the active system (lines) with the same value of $(\lambda-\lamc)$. We have also explored this behavior for other values of $f_0$ and $\tau_p$ that enter the active non-stationary MCT via $\Delta(t)$ in Eqs. (\ref{correq1}) and (\ref{responseeq1}) (see SM Fig. S2).   We find that the distance from the corresponding critical points, that is $(\lambda-\lamc)$, always governs the behavior of $t_r(t_w)$.
	
	\begin{figure}
		\includegraphics[width=8cm]{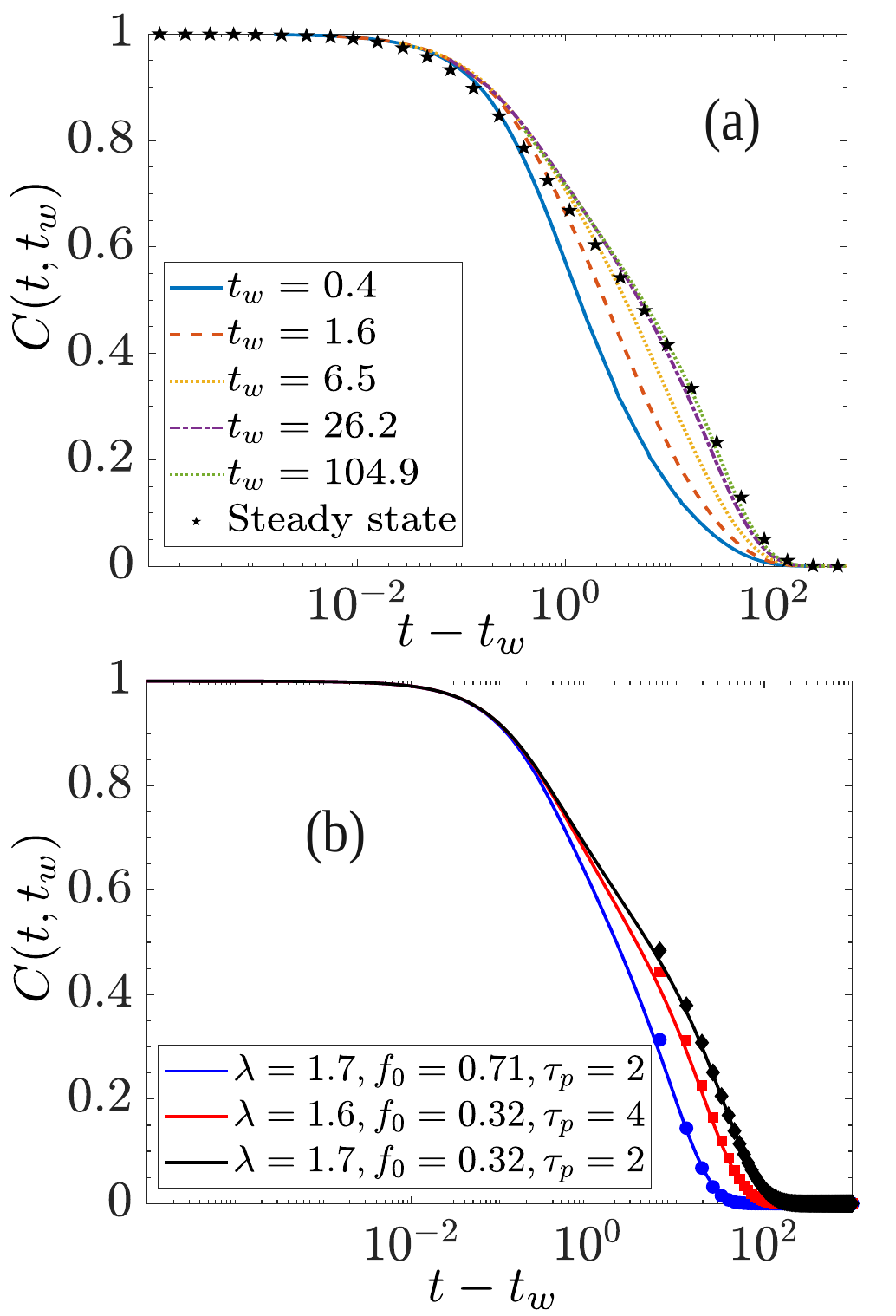}
		\caption{Comparison of Aging and Steady-State Solutions. (a) When the quench value of $\lambda$ is such that $\lambda<\lamc$, the non-stationary state evolves towards a stationary state, and $t_r$ saturates after some $t_w$. This stationary state agrees with the steady-state MCT for the active glasses (shown by the symbols). We have taken $f_0=0.32$, $\tau_p=2$ and quenched the system to $\lambda=1.7$.
			(b) Comparison of the stationary solutions (the symbols) of the generic MCT when $t_w$ is large, with the steady-state active MCT (lines) for quench to various activity values and $\lambda$. The symbols and lines with the same color represent identical parameter sets for the generic MCT and the steady-state active MCT.
		}
		\label{fig3}
	\end{figure}

We next demonstrate that the same result also holds for the aging dynamics when the quench is in the liquid regime. To prove this, we quench a passive system for two values of $\lambda$ and follow the evolution of $t_r(t_w)$ [Fig. \ref{aging}b]. Concurrently, we take an active system with $f_0^2=0.12$ and $\tau_p=1.0$ and quench it to two values of $\lambda$ (Fig. \ref{aging}b). Note that even for these values of activity parameters, \textcolor{black}{we have $\delta \lambda_c \simeq 0.2$; we chose the quench parameters such that for each passive system, there is a corresponding active system such that they will have the same values of $(\lammct-\lambda)$ and $(\lamc-\lambda)$, respectively.} For systems with identical distances of their quench from their respective critical points, $t_r$ as a function of $t_w$ overlaps. In addition, since the quench is below the MCT critical points, we expect that $t_r$ will saturate at large $t_w$; the results in Fig. \ref{aging}(b) are consistent. 

Thus, active aging has similarities with the aging dynamics in passive systems, and the distance from the critical point governs the active aging dynamics. We will further show in Sec. \ref{comp_pred} that the activity-modification of the MCT critical points has further consequences for the aging dynamics. However, before that, we discuss the approach to the steady state when the quench is in the liquid regime.

\subsection{Evolution towards the steady state}
\label{evol_SS}
We now show that when we quench the system in the liquid phase, it evolves towards a stationary state and $t_r$ saturates after some $t_w$. In addition, the stationary state agrees with the steady-state active MCT derived in Ref. \cite{nandi2017nonequilibrium}. This agreement has further consequences for the active MCT for the following reason. 
MCT describes the glass transition as a critical phenomenon with a transition at $\lammct$. \textcolor{black}{This} is a nontrivial problem even for equilibrium MCT. One way to derive the equilibrium MCT is to take the $t_w\to\infty$ limit of the generic nonequilibrium non-stationary MCT that describes a system even under aging \cite{bouchaud1996}. If the approximations are reasonable, the theory should agree with the final form via other approaches, such as the projection operator formalism \cite{gotze2009complex,reichman2005mode} or the one starting with Newton's equations \cite{zaccarelli2001}. Two forms agree in the liquid state, but not in the non-ergodic regime. The former requires $C(t,t_w)\to0$ as $t_w\to\infty$; however, the resulting theory predicts a nonergodicity transition. 
In addition, unlike the equilibrium MCT, various approaches to derive the active MCT lead to slightly different variants of the theory \cite{liluashvili2017,szamel2015,feng2017,nandi2017nonequilibrium,sadhukhan2024perspective}. This shows the complex nature of the system and that various approximations in the derivation become even more obscure in the presence of activity. A demonstration that the long-time limit of the non-stationary MCT agrees with the steady state variant is therefore quintessential for active systems.

We first show that when the quench is below $\lamc$, i.e., in the liquid state, the non-stationary state evolves towards a stationary state. Figure \ref{fig3}(a) shows the evolution of the non-stationary state towards the steady state as $t_w$ increases for $f_0=0.32$, $\tau_p=2$, and the quench value of $\lambda=1.7$. The solutions overlap with each other beyond $t_w=26.2$. The corresponding $t_r$ will grow as $t_w$ increases at small $t_w$, and then saturate. Figure \ref{aging}(b) shows the saturation of $t_r(t_w)$ when the quench is in the liquid phase. We have also solved the steady state active MCT of Ref. \cite{nandi2017nonequilibrium} for the same set of $f_0$, $\tau_p$, and $\lambda$ and show the correlation function $C(t)$ for comparison: it agrees with the stationary $C(t,t_w)$ in the limit of $t_w\to\infty$. Note that the tiny difference in the small time, as discussed in the SM, is due to different accuracies of the two implementations. Figure \ref{fig3}(b) shows the comparison of the saturated $C(t,t_w)$ with the steady state MCT result for several other parameters. These results prove that the generic theory agrees with the steady state MCT for active systems. This agreement confirms that the approximations involved in the active MCT of Ref. \cite{nandi2017nonequilibrium} are comparable to those of the equilibrium theory. A similar comparison for the other approaches of active MCT will be illuminating to reveal the nature of the mode-coupling approximations involved in these derivations.

\begin{figure}
\includegraphics[width=8.8cm]{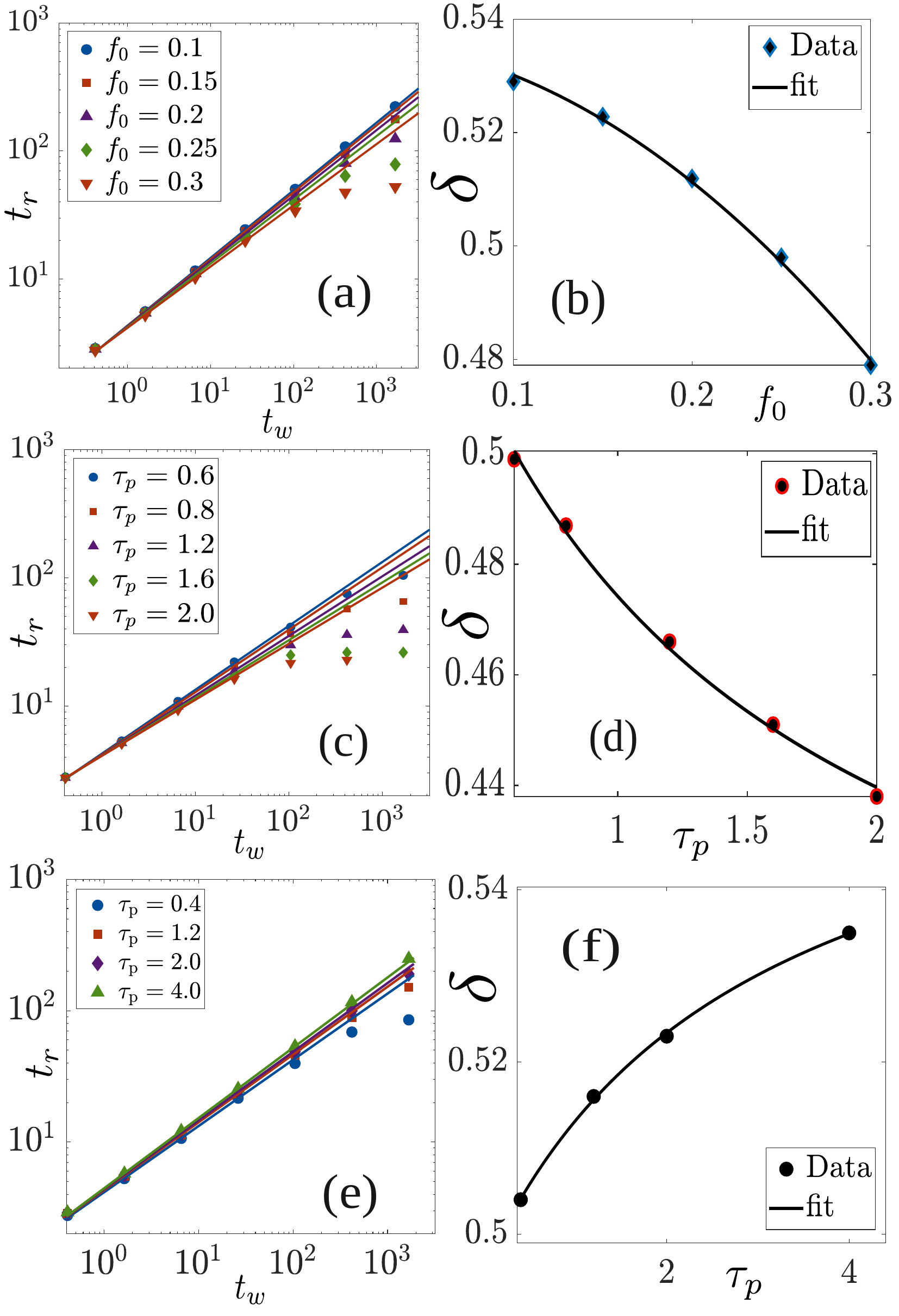}
\caption{Predictions of the theory. (a) The active system ages faster, leading to smaller $t_r$ at the same $t_w$ as $f_0$ increases. Lines are fits of the early-$t_w$ data with a power law $t_r \sim t_w^\delta$, and symbols are numerical solutions. We have taken $\tau_p = 1$ and quenched at $\lambda = 1.9$. (b) Plot of $\delta$ corresponding to the data in (a) as a function of $f_0$ (symbols). The line is a fit with the function $f(x)=A-bx^2$ with $A= 0.54$ and $b= 0.63$. (c)$t_r$ as a function of $t_w$ for a quench at $\lambda = 1.999$ with constant $f_0 = 0.4$ and various $\tau_p$. Lines are the fits of the data to the early $t_w$-regime with $t_r\sim t_w^\delta$. (d) Plot of $\delta$ as a function of $\tau_p$ for the data in (c). Line is fit with a function $f(x)=A-bx/(1+Cx)$ with $A$ = 0.58, $b=0.21$, and $C=1.05$. (e) For the AOUP system, we show $t_r$ as a function of $t_w$ for a quench to $\lambda = 1.99$ at constant $f_0 = 0.3$ and various $\tau_p$. Lines are fits of $t_r\sim t_w^\delta$ to the early $t_w$ data. (f) For the AOUP system, $\delta$ increases for larger $\tau_p$ for the data in (e). Line is fit with $f(x)=A-b/(1+cx)$ with $A = 0.56$, $b=0.07$, and $c=0.35$.}
\label{predictions}
\end{figure}

\subsection{Predictions of the theory and comparison with existing simulations}
\label{comp_pred}
We now spell out further predictions of the theory on the aging dynamics of active glasses and compare them with existing simulation results whenever possible. Within MCT, the glassy properties are governed by the MCT critical point, but the transition itself is avoided in simulations and experiments. Above the transition, other mechanisms that are external to MCT take over. Therefore, to compare with simulations, we quench the system in the liquid regime, but close to $\lamc$. We show that the modification of the critical point due to activity has further significance for the aging dynamics in active glasses. 

We show $t_r$ as a function of $t_w$ in Fig. \ref{predictions} for different activity parameters and models of activity. Figure \ref{predictions}(a) shows the evolution of $t_r$ as a function of $t_w$ when we quench the system to $\lambda=1.9$ and a fixed $\tau_p=1$, and with various $f_0$. We fit the low-$t_w$ part of the data with a power law form: $t_r\sim t_w^\delta$ \cite{kob1997aging, warren2013quench,klongvessa2022aging, nandi2012glassy, mandal2020multiple, janzen2022aging}. The lines in Fig. \ref{predictions}(a) show the fits with the data (symbols). Figure \ref{predictions}(b) shows that $\delta$ decreases as $f_0$ increases. In fact, the variation of $\delta$ with activity also depends on the distance of the quench from $\lamc$. As we have argued in the SM, Sec. IV, the aging exponent in the presence activity for the ABP system will vary as
\begin{equation}\label{delta}
\delta=A-Bf_0^2\tau_p/(1+C\tau_p),
\end{equation}
where $A$, $B$, and $C$ are constants. For constant $\tau_p$, we can write Eq. (\ref{delta}) as $\delta=A-bf_0^2$. The line in Fig. \ref{predictions}(b) is a fit with this form; it agrees well with the numerical solution. This prediction agrees well with the simulation results of Ref. \cite{mandal2020multiple}. We have also extracted $\delta$ with varying $f_0$ from the data of Ref. \cite{janzen2022aging}, and Eq. (\ref{delta}) agrees well with these simulation data as well (see SM Fig. S3). Figure \ref{predictions}(c) shows the evolution of $t_r$ with $t_w$ for a quench $\lambda=1.999$ with a fixed $f_0=0.4$ and various $\tau_p$. When $f_0$ is fixed, we can write Eq. (\ref{delta}) as $\delta=A-b\tau_p/(1+C\tau_p)$. We show the fit of this form with $\delta$ as a function of $\tau_p$ in Fig. \ref{predictions}(d) by the line along with the numerical solution (symbols). The constants in Eq. (\ref{delta}) will be identical for different active systems only when the quenches are the same.

For comparison, we also studied the aging behavior in the AOUP system. We know that the glassy behaviors for the two models of activity are similar when we vary $f_0$ \cite{nandi2017nonequilibrium,nandi2018random}. However, the behaviors differ when we vary $\tau_p$. We find that the aging dynamics also has a similar trend. Therefore, we present the results for varying $\tau_p$ alone. Figure \ref{predictions}(e) shows the evolution of $t_r$ with $t_w$ for a quench $\lambda=1.99$ with fixed $f_0=0.3$ and varying $\tau_p$; the trend is opposite to that in the ABP model (Fig. \ref{predictions}c and d). Figure \ref{predictions}(f) shows the values of $\delta$ as a function of $\tau_p$ for the data in Fig. \ref{predictions}(e). For the AOUP model, we will have $\delta=A-Bf_0^2/(1+C\tau_p)$ [see SM, Sec. IV]. At constant $f_0$, we can write $\delta=A-b/(1+c\tau_p)$. The line in Fig. \ref{predictions}(f) shows the fit with this analytical form; it agrees remarkably well. Thus, contrary to the ABP system, the aging dynamics in the AOUP system becomes slower as $\tau_p$ increases.

\section{Discussion}
\label{disc}
We have obtained the nonstationary mode-coupling theory for the aging dynamics in active glasses. The primary technical challenge for progress in this direction was the absence of a suitable numerical algorithm to solve the nonstationary active MCT. We have now developed such an algorithm and show that the aging properties are governed by the distance of the quench from the critical point, $\lamc$. Similar to the steady state dynamics, the aging dynamics for the two models of activity, ABP and AOUP, are opposite when we vary $\tau_p$. For the ABP system, $\lamc$ increases as $\tau_p$ grows. Therefore, for the same quench in $T$, represented by $\lambda$ within the schematic theory, the active glass ages faster than passive glasses, and the power law exponent $\delta$ decreases. By contrast, as $\lamc$ decreases for larger $\tau_p$ in the AOUP system, the aging dynamics becomes slower, and $\delta$ increases. The predictions remain to be tested in simulations. On the other hand, both models are equivalent when we vary $f_0$. In that case, the active system ages faster, and $\delta$ decreases. This result is consistent with existing simulation results on ABP systems \cite{mandal2020multiple,janzen2022aging}.

An interesting direction for future works will be to extend the theory for the $\tau_p\to\infty$ limit. Activity naturally leads to the separation of time scales via $\tau_p$; setting this time scale to the extreme limit provides the so-called ``extreme active matter" \cite{mandal2020_natcomm,keta2023intermittent,szamel2024, mandal2022random}. Simulations have shown that this limit can have further intriguing aging dynamics, for example, leading to multiple decay of the correlation functions for the ABP system \cite{mandal2020multiple}. However, this regime for the AOUP system might be different, as in addition to the active force directions being quenched, the force magnitude also tends to zero. For the steady-state dynamics in AOUPs, the nature of the critical point changes with varying $\tau_p$, from glass-like at small $\tau_p$ to jamming-like at large $\tau_p$ \cite{pareek2025}. How this change in the critical property affects the aging dynamics remains unknown.

The current work has crucial significance for the active MCT. There are several possible routes to derive MCT; they lead to the same final form for equilibrium systems. However, additional approximations are necessary to derive the steady-state active MCT. The existing versions of active MCT for the steady state vary  \cite{feng2017,szamel2015,flenner2016,berthier2013,liluashvili2017,nandi2017nonequilibrium,debets2022}, and the detailed form depends on the specific method of derivation \cite{sadhukhan2024perspective}. This difference illustrates that the additional approximations of the theory due to activity alone lead to further complications. By contrast, the theory for the non-stationary state is more generic. We have shown that the steady-state active MCT obtained via the field-theoretic route agrees with the non-stationary generic MCT after it has evolved to the stationary state when quenched to the liquid regime. A more detailed comparison of the field-theoretic derivation with other approaches will be instructive.

Aging phenomena in biological systems are wide-ranging and have far-reaching consequences. The parameters can also be quite different from typical particulate systems. For example, maturation of junction proteins, spatiotemporally coordinated cell divisions and apoptosis, sudden change in nutrient concentration and physical conditions, etc. We can include some of these features in the vertex-based models of confluent tissues \cite{farhadifar2007,fletcher2014,barton2017}. Aging in these latter systems has further complexity, possibly due to the long-range nature of the models, and will be discussed elsewhere. The particulate models of active matter are convenient starting points for tissue models and capture several aspects of the former despite their simplistic approximations. For example, the sub-to-super Arrhenius transition and the modification of the glass transition point are similar in both classes of models \cite{sadhukhan2024,pareek2025}. Our work provides a theoretical framework for the aging dynamics in the presence of activity and is a vital step toward understanding how aging dynamics is related to the properties of biological systems.

\section{Acknowledgments}
We acknowledge the support of the Department of Atomic Energy, Government of India, under Project Identification No. RTI 4007.


%



\pagebreak
\newpage
\widetext

\begin{center}
	\textbf{\large Supplementary Material: Mode-Coupling Theory for aging in active Glasses: relaxation dynamics and evolution towards steady state}
	\vspace{0.5cm}
	
	Soumitra Kolya$^{1}$, Nir S. Gov$^{2}$, and Saroj Kumar Nandi$^{1}$
	
	\vspace{0.2cm}
	
	{\small
		$^{1}$ Tata Institute of Fundamental Research, Gopanpally Village, Hyderabad - 500046, India \\
		$^{2}$ Department of Chemical and Biological Physics, Weizmann Institute of Science, Rehovot 7610001, Israel
	}
	
	\vspace{0.2cm}

\end{center}


\setcounter{table}{0}
\setcounter{figure}{0}
\setcounter{section}{0}
\setcounter{equation}{0}
\renewcommand{\thefigure}{S\arabic{figure}}
\renewcommand{\thetable}{S\arabic{table}}
\renewcommand{\theequation}{S\arabic{equation}}
\renewcommand{\thepage}{S\arabic{page}}
\renewcommand{\thesubsection}{S\arabic{subsection}}
\renewcommand{\bibnumfmt}[1]{[S#1]}
\renewcommand{\citenumfont}[1]{S#1}
\renewcommand{\d}{\mathrm{d}}
\renewcommand{\r}{\textcolor{red}}
\renewcommand{\u}{\mathbf{u}^{\text{ext}}}
\renewcommand{\theequation}{\arabic{equation}}
\renewcommand{\thefigure}{\arabic{figure}}
\renewcommand{\thesection}{\Roman{section}}
\renewcommand{\a}{\alpha}
\renewcommand{\b}{\beta}
\renewcommand{\d}{\mathrm{d}}
\renewcommand{\t}{\tau}
\renewcommand{\S}{\Sigma}
\renewcommand{\l}{\lambda}
\renewcommand{\thefigure}{S\arabic{figure}}
\renewcommand{\thetable}{S\arabic{table}}
\renewcommand{\theequation}{S\arabic{equation}}
\renewcommand{\thepage}{S\arabic{page}}
\renewcommand{\thesubsection}{S\arabic{subsection}}

\section{Algorithm for solving the non-stationary mode-coupling theory for active glasses}

The schematic form of the generic nonequilibrium non-stationary mode-coupling theory (MCT) equations, as derived in the main text, applicable for an active system undergoing aging is

\begin{align}
	\frac{\partial C(t,t_w)}{\partial t} &= -\mu(t) C(t,t_w) + 2T R(t_w,t)+\int_0^{t_w} \mathrm{d}s\, \mathcal{D}(t,s) R(t_w,s) + \int_0^t \mathrm{d}s\, \Sigma(t,s) C(s,t_w), \label{correq1} \\
	\frac{\partial R(t,t_w)}{\partial t} &= -\mu(t) R(t,t_w) + \delta(t - t_w) + \int_{t_w}^t \mathrm{d}s\, \Sigma(t,s) R(s,t_w), \label{responseeq1} \\
	\mu(t) &= T + \int_0^t \mathrm{d}s\, [\mathcal{D}(t,s) R(t,s) + \Sigma(t,s) C(t,s)] \label{def_mu},
\end{align}
where $\mathcal{D}(t,s) = 2\lambda C^2(t,s) + \Delta(t-s)$ and \(\Sigma(t,s) = 4\lambda C(t,s) R(t,s)\). Note that within the schematic form, \(\lambda\) contains the information of changing parameters, such as $T$ or density. As these parameters vary, the static structure factor and the direct correlation functions change. Their values at a particular wavevector and the vertex function lead to the $\lambda$. Therefore, we can set $T=1$ and use $\lambda$ as the control parameter.

We will write the theory of the correlation function and the integrated response function. $F(t,t_w)$, defined as
\begin{equation}
	F(t,t_w) = -\int_{t_w}^t R(t,s)\, \mathrm{d}s.
\end{equation}
Such a representation is advantageous for the numerical integration since the fluctuation in $F(t,t_w)$ is less than that in $R(t,t_w)$. With a little bit of straightforward algebra, we can write Eq. (\ref{responseeq1}) as
\begin{align}\label{varchange2}
	\frac{\partial F(t,t_w)}{\partial t} = -1 - \mu(t) F(t,t_w)
	+ \int_{t_w}^t \mathrm{d}s\, \Sigma(t,s) F(s,t_w).
\end{align}
We use the definitions of $\mathcal{D}(t,s)$ and $\Sigma(t,s)$, given above and explicitly write the equations for $C(t,t_w)$ and $F(t,t_w)$ as
\begin{subequations}
	\label{correq2}
	\begin{align}
		\frac{\partial C(t,t_w)}{\partial t} =& -\mu(t) C(t,t_w) 
		+ 2\lambda \int_0^{t_w} \mathrm{d}s \, C^2(t,s) \frac{\partial F(t_w,s)}{\partial s} 
		\quad + 4\lambda \int_0^t \mathrm{d}s \, C(t,s) \frac{\partial F(t,s)}{\partial s} C(s,t_w) \\
		&+ \int_0^{t_w} \Delta(t-s) \frac{\partial F(t_w,s)}{\partial s} \, \mathrm{d}s \\
		\frac{\partial F(t,t_w)}{\partial t} =& -1 - \mu(t) F(t,t_w) 
		+ 4\lambda \int_{t_w}^t \mathrm{d}s \, C(t,s) \frac{\partial F(t,s)}{\partial s} F(s,t_w) \\
		\text{and,}\,\,\,	\mu(t) =& T + \int_0^t \mathrm{d}s \left[
		\left( 2\lambda C^2(t,s) + \Delta(t-s) + 4\lambda C^2(t,s) \right)
		\frac{\partial F(t,s)}{\partial s} \right].
	\end{align}
\end{subequations}
The correlation function for glassy systems has the following generic properties. It decays very fast at small times and extremely slow at large times. To capture this decay property within the numerical algorithm, one should consider an adaptive step size that must be very small at short times and becomes progressively larger as time grows. This is the first challenge for the numerical solution for the aging dynamics as this property must hold for both $t$ and $t_w$, and the computation time requirement becomes enormous. One can save some computation time by solving the equations in the time domain of $(t,\tau=t-t_w)$ as the time domain now becomes half of the original requirement \cite{nandi2012glassy,nandi2016glass,kimlatz}. Therefore, write the theory in the $(t,\tau)$ domain, the equations for $F(t, \tau)$ and $C(t, \tau)$  become
\begin{subequations}\label{disceq}
	\begin{align}
		\left(\frac{\partial}{\partial t} + \frac{\partial}{\partial \tau}\right) F(t, \tau) &= -1 - \mu(t) F(t, \tau) - 4\lambda \int_0^\tau \frac{\partial F(t, s)}{\partial s} C(t, s) F(t - s, \tau - s) \, ds \label{respdisc}\\
		\left(\frac{\partial}{\partial t}+\frac{\partial}{\partial \tau}\right)C(t,\tau) &= -\mu(t) C(t,\tau)+2\lambda\int_\tau^t \frac{\partial C^2(t,s)}{\partial s} F(t-\tau,s-\tau) d s 
		-2\lambda C^2(t,t)F(t-\tau,t-\tau) \nonumber\\
		&-4\lambda\int_\tau^t C(t,s)\frac{\partial F(t,s)}{\partial s}C(t-\tau,s-\tau)\d s 
		-4\lambda\int_0^\tau C(t,s)\frac{\partial F(t,s)}{\partial s}C(t-s,\tau-s)d s \nonumber\\
		&-\Delta(t) F(t-\tau, t-\tau) + \int_\tau^t \frac{\Delta(s)}{\partial s} F(t-\tau,s-\tau) d s \label{corrdisc},
	\end{align}
\end{subequations}
with
\begin{equation}
	\mu(t) = T - 6\lambda \int_0^t C^2(t, s) \frac{\partial F(t, s)}{\partial s} ds - \int_0^t \Delta(s) \frac{\partial F(t, s)}{\partial s} \, ds = T - 6\lambda \epsilon(t) - p(t). \label{mueq}
\end{equation}
We are now ready to discretize these equations for the numerical solution. However, compared to the aging dynamics equations of passive systems \cite{nandi2012glassy}, Eqs. (\ref{disceq}-\ref{mueq}) have additional difficulties due to the activity terms that make the solution even more challenging. We will discuss this later.

\subsection{Discretization of the equations of motion}

The discretization procedure is a bit involved due to the adaptive grid size and two-dimensional nature of the problem. The primary goal is to transform the integrals such that we can evaluate the derivatives appearing in the integrand in the long-time regime, where the functions have relatively smooth variation. We can achieve this task by folding the integrals, which leads to several time points in the discretized form. We must keep track of these time points, and a specific notation becomes helpful. We follow the same notation introduced by Kim and Latz \cite{daniel2000,kimlatz} for the numerical algorithm of the aging dynamics in passive systems. We discretize the time grid into $i$ and define the functions $i(t)$ and $t(i)$; the first gives the discrete time point $i$ for a given continuous time $t$, and the second provides the opposite. We define the step size $h(i)$ and double it every $N_s$ step. For a choice of $h(i)$, it is easy to define the functions $i(t)$ and $t(i)$. For the time derivatives, we define them at the current time as follows
\begin{equation}
	\frac{\partial f(s)}{\partial s}=\frac{f(i_s)-f(i_s-1)}{h(i_s)},
\end{equation} 
and we take the functions as averages with the next time point: $g(s)=[g(i_s)+g(i_s-1)]/2$. This strategy is for enhancing numerical accuracy. We first discretize the equation for the integrated response function. We can write down Eq. (\ref{respdisc}) in the discretized notation as

\begin{equation}
	\frac{F(i_t,i_\tau) - F'}{h(i_\tau)} = -1 - \mu(i_t) F(i_t, i_\tau) - 4\lambda \cdot \text{IntegralF}
\end{equation}
where we have written $F'$ as
\[
F' = F(i_t - 1, i_\tau - 1) + \left( F(i_t, i_\tau - 1) - F(i_t - 1, i_\tau - 1) \right) \left(1-\frac{h(i_\tau)}{t(i_t) - t(i_t - 1)}\right)
\]
and,
\begin{equation}
	\begin{aligned}
		\text{IntegralF} = &\; \frac{1}{2}[F(i_t, 1) - F(i_t, 0)] \cdot C(i_t, 0) \cdot F(i_t, i_\tau) \\
		& + \frac{1}{2}[F(i_t, i_h) - F(i_t, i_h - 1)] C(i_t, i_h) F(i_{1h}, i_{2h}) \\
		& + \frac{1}{2} \sum_{i_s=1}^{i_h-1} [F(i_t, i_s + 1) - F(i_t, i_s - 1)] C(i_t, i_s) F(i_1, i_2) \\
		& - \frac{1}{2}[F(i_t, i_{2p0}) - F(i_t, i_\tau)] C(i_t, i_\tau) F(i_3, 0) \\
		& - \frac{1}{2}[F(i_t, i_{2h}) - F(i_t, i_{2mh})] C(i_t, i_{2h}) F(i_{1ph}, i_h) \\
		& - \frac{1}{2} \sum_{i_s=1}^{i_h-1} [F(i_t, i_{2p}) - F(i_t, i_{2m})] C(i_t, i_2) F(i_5, i_s)
	\end{aligned}
\end{equation}
with the definitions of various indices as follows:
\begin{align*}
	&	i_h = i(t(i_\tau/2)), &&   i_{2mh} = i(t(i_\tau) - t(i_h - 1)), \\
	&	i_{1h} = i(t(i_t) - t(i_h)),  && i_{1ph} = i(t(i_t) - t(i_\tau) + t(i_h)), \\
	&	i_{2h} = i(t(i_\tau) - t(i_h)),  && i_{2p} = i(t(i_\tau) - t(i_s + 1)) \\
	&	i_1 = i(t(i_t) - t(i_s)),  && i_{2m} = i(t(i_\tau) - t(i_s - 1)), \\
	&	i_2 = i(t(i_\tau) - t(i_s)), 	&& i_5 = i(t(i_t) - t(i_\tau) + t(i_s)), \\
	&	i_3 = i(t(i_t) - t(i_\tau)),   && i_{2p0} = i(t(i_\tau) - t(1))
\end{align*}

We will use this same notation for the other terms as well and provide the additional indices below. We now write the descritized version of Eq. (\ref{corrdisc}):
\begin{equation}
	\frac{C(i_t, i_\tau) - C'}{h(i_\tau)} = -\mu(i_t) C(i_t, i_\tau) - 2\lambda C^2(i_t, i_\tau) F(i_3, i_3) + 2\lambda \cdot \text{IntegralC3} - 4\lambda(\text{IntegralC1} + \text{IntegralC2}) + \Pi(i_t,i_{\tau})
	\label{corr_function}
\end{equation}
where, equation of $C'$ is
\begin{equation}
	C' = C(i_t - 1, i_\tau - 1) + \left( C(i_t, i_\tau - 1) - C(i_t - 1, i_\tau - 1) \right) \left(1-\frac{h(i_\tau)}{t(i_t) - t(i_t - 1)}\right)
\end{equation}
and the other terms are as follows:

\begin{equation}
	\begin{aligned}
		\text{IntegralC1} = &\; \frac{1}{2}[F(i_t, 1) - F(i_t, 0)] C(i_t, 0) C(i_t, i_\tau) \\
		& + \frac{1}{2}[F(i_t, i_h) - F(i_t, i_h - 1)] C(i_t, i_h) C(i_{1h}, i_{2h}) \\
		& + \frac{1}{2} \sum_{i_s=1}^{i_h-1} [F(i_t, i_s + 1) - F(i_t, i_s - 1)] C(i_t, i_s) C(i_1, i_2) \\
		& - \frac{1}{2}[F(i_t, i_{2p0}) - F(i_t, i_\tau)] C(i_t, i_\tau) C(i_3, 0) \\
		& - \frac{1}{2}[F(i_t, i_{2h}) - F(i_t, i_{2mh})] C(i_t, i_{2h}) C(i_{1ph}, i_h) \\
		& - \frac{1}{2} \sum_{i_s=1}^{i_h-1} [F(i_t, i_{2p}) - F(i_t, i_{2m})] C(i_t, i_2) C(i_5, i_s)
	\end{aligned}
\end{equation}

\begin{equation}
	\begin{aligned}
		\text{IntegralC2} = &\; \frac{1}{2}[F(i_t, i_{2m0}) - F(i_t, i_\tau)] C(i_t, i_\tau) C(i_3, 0) \\
		& + \frac{1}{2}[F(i_t, i_{63}) - F(i_t, i_{6m3})] C(i_t, i_{63}) C(i_3, i_3) \\
		& + \frac{1}{2} \sum_{i_s=1}^{i_3-1} [F(i_t, i_{6p}) - F(i_t, i_{6m})] C(i_t, i_6) C(i_3, i_s)
	\end{aligned}
\end{equation}

\begin{equation}
	\begin{aligned}
		\text{IntegralC3} = &\; \frac{1}{2}[C^2(i_t, i_{2m0}) - C^2(i_t, i_\tau)] F(i_3, 0) \\
		& + \frac{1}{2}[C^2(i_t, i_{63}) - C^2(i_t, i_{6m3})] F(i_3, i_3) \\
		& + \frac{1}{2} \sum_{i_s=1}^{i_3-1} [C^2(i_t, i_{6p}) - C^2(i_t, i_{6m})] F(i_3, i_s)
	\end{aligned}
\end{equation}

The terms involving the active forces are obtained via iteration as we discuss below. $\Pi(i_t, i_\tau)$ has two parts. The first part is similar to that used for the second term in RHS of Eq.~\ref{corr_function}.\\
The second part of $\Pi(i_t,i_{\tau})$ is as follows:
\begin{equation}
	\begin{aligned}
		\text{Second term in $\Pi(i_t,i_{\tau})$} = &\; \frac{1}{2}[\Delta(i_{2m0}) - \Delta(i_\tau)] F(i_3, 0) \\
		& + \frac{1}{2}[\Delta(i_{63}) - \Delta(i_{6m3})] F(i_3, i_3) \\
		& + \frac{1}{2} \sum_{i_s=1}^{i_3-1} [\Delta(i_{6p}) - \Delta(i_{6m})] F(i_3, i_s)
	\end{aligned}
\end{equation}

The term $\mu(t)$ has two parts involving the integrals, we have designated them as $\epsilon(t)$ and $p(t)$ in Eq. (\ref{mueq}). The expression for $\epsilon(i_t)$ is
\begin{equation}
	\begin{aligned}
		\epsilon(i_t) = &\; \frac{1}{2}[F(i_t, 1) - F(i_t, 0)] C^2(i_t, 0) \\
		& + \frac{1}{2}[F(i_t, i_t) - F(i_t, i_t - 1)] C^2(i_t, i_t) \\
		& + \frac{1}{2} \sum_{i_s=1}^{i_t-1} [F(i_t, i_s + 1) - F(i_t, i_s - 1)] C^2(i_t, i_s).
	\end{aligned}
\end{equation}
The other term, $p(t)$, involves activity. We discretize this term as follows:
\begin{equation}
	\begin{aligned}
		p(i_t) = &\; \frac{1}{2}[F(i_t, 1) - F(i_t, 0)] \Delta(0) \\
		& + \frac{1}{2}[F(i_t, i_t) - F(i_t, i_t - 1)] \Delta(i_t) \\
		& + \frac{1}{2} \sum_{i_s=1}^{i_t-1} [F(i_t, i_s + 1) - F(i_t, i_s - 1)] \Delta(i_s)
	\end{aligned}
\end{equation}

The additional indices, used in the discretization of the correlation function equation and $\mu(t)$ are
\[
\begin{aligned}
	&	i_{2m0} = i(t(i_\tau) + t(1)), && i_{6m3} = i(t(i_\tau) + t(i_3 - 1)) \\
	&	i_6 = i(t(i_s) + t(i_\tau)), && i_{63} = i(t(i_3) + t(i_\tau)) \\
	&	i_{6p} = i(t(i_\tau) + t(i_s + 1)), \hspace{2cm}  &&i_{6m} = i(t(i_\tau) + t(i_s - 1)) . 
\end{aligned}
\]

We now provide the algorithm to solve the aging equations for active systems.

\subsection{Algorithm to solve the non-stationary MCT for active aging dynamics}
A close look at the equations of the active aging theory reveals that we need $C(t,\tau)$ and $F(t,\tau)$ at all times to obtain the terms involving activity. Therefore, we use a self-consistent iterative approach. We first solve the equations assuming activity is zero and obtain $C$ and $F$. We use these solutions, evaluate the activity-containing terms, and then solve for $C$ and $F$ again. We repeat this until the values for the terms containing activity saturate.

\begin{figure}
	\centering
	\includegraphics[width=9.6cm,height=6.6cm]{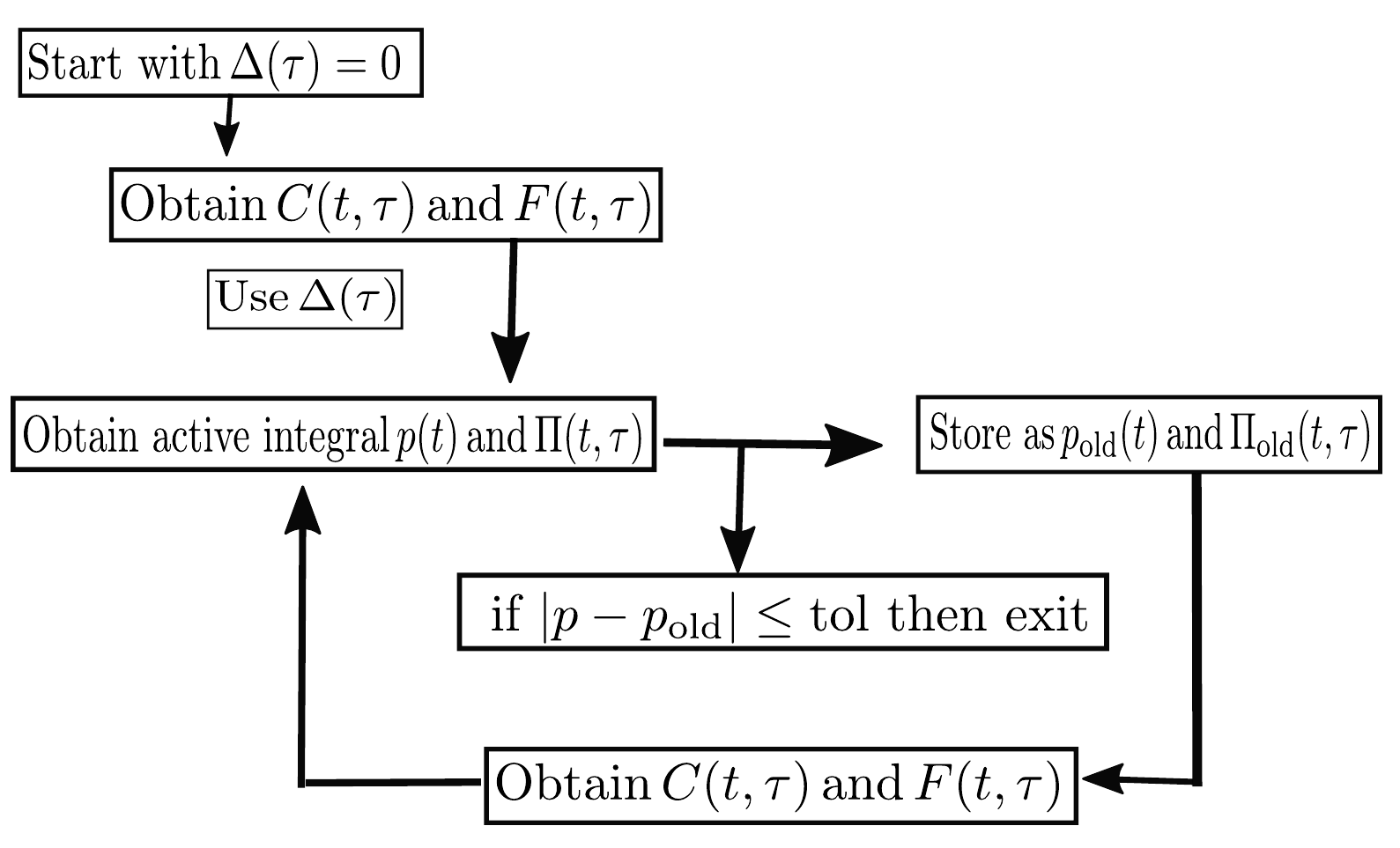}
	\caption{Schematic representation of the numerical algorithm used to solve the non-stationary mode-coupling theory equations applicable for active aging dynamics.}
	\label{schematic}
\end{figure}

Although this process sounds straightforward, it is nontrivial in practice as the aging solution, even for the passive glasses, is time-consuming (order of several hours for a reasonable waiting time-dependent data). Therefore, several fine-tuning of the parameters is necessary so that the solution converges within a reasonable time. One such fine-tuning is the value of the initial time step, which cannot be arbitrarily small, and the accuracy of the aging solutions will be lower compared to that of the steady-state solution. This shows up when we compare the stationary solution of the aging theory with the steady-state solution (for example, Fig. 3 in the main text). We now sketch out the steps to numerically solve the discretized theory. Figure \ref{schematic} shows a schematic flow chart of the algorithm.

\begin{itemize}
	\item {Initialization:} 
	\begin{itemize}
		\item As discussed above, we start the solution by setting $\Delta(\tau) = 0$.
		\item Since we have written the theory in the $(t,\tau)$ parameterization, the initial conditions are trivial, and given as follows: for all \( i_s \in \{0, \ldots, N\} \), $C(i_s, 0) = 1, \quad F(i_s, 0) = 0$.
		
	\end{itemize}
	
	\item The MCT solution at each time step are also solved iteratively. For $i_t = 1 \ldots N$, we define some guess values,
	\begin{equation}
		C_\text{new}(i_s) = C(i_t-1, i_s-1), \quad 
		F_\text{new}(i_s) = F(i_t-1, i_s-1) \quad \forall i_s \in \{1, \ldots, i_t\}
	\end{equation}
	and use these guess values for the new solution
	\[
	C(i_t, i_s) = C_\text{new}(i_s), \quad 
	F(i_t, i_s) = F_\text{new}(i_s) \quad \forall i_s \in \{1, \ldots, i_t\}.
	\]
	
	We then solve the equations of motion and evaluate the ``correct'' solutions. We store these solutions to the corresponding variables
	\[
	C(i_t, i_s) = C_\text{new}(i_s), \quad 
	F(i_t, i_s) = F_\text{new}(i_s) \quad \forall i_s \in \{1, \ldots, i_t\}.
	\]
	We continue this process till we obtain the desired accuracy:
	\[
	\text{Norm}_{i_s \in \{1, \ldots, i_t\}} \left( C_\text{new}(i_s) - C(i_t, i_s), \, F_\text{new}(is) - F(i_t, i_s) \right) < \text{desired accuracy}
	\]
	and then save the converged solutions for $C(i_t, i_s)$ and $F(i_t, i_s)$ at each time step $i_t$.
	
	\item {Evaluate the terms containing activity:}
	We next use the non-zero value of activity, depending on the model we are using. For example, the ABP activity is 
	\[
	\Delta(t) = f_0^2 \exp(-t/\tau_p).
	\]
	We now evaluate the terms involving the activity: \( p(t) \) and \( \Pi(t, \tau) \). Store them as: $
	p_{\text{old}}(i_t) = p(i_t), \quad \\
	\Pi_{\text{old}}(i_t, i_s) = \Pi(i_t, i_s)$
	for the use in the next iteration.
	
	\item We next reevaluate the correlation and response functions using the non-zero $\Delta(t)$.
	
	\item
	We then again compute \( p(t) \) and \( \Pi(i_t, i_s) \) using the newly updated correlation and response functions. We continue the entire procedure until we obtain
	$|p - p_{\text{old}}| < \text{desired accuracy}$
	where \( p = \sum_t p(t) \) denotes the total active contribution.
	
	Once the calculation converges, We store the final solution for \( C(i_t, i_s) \) and \( F(i_t, i_s) \). These are the aging solutions for the active system.
\end{itemize}

\section{Distance from criticality governs the behaviour of $t_r(t_w)$}

\begin{figure}[h]
	\centering
	\includegraphics[width=15cm,height=6.0cm]{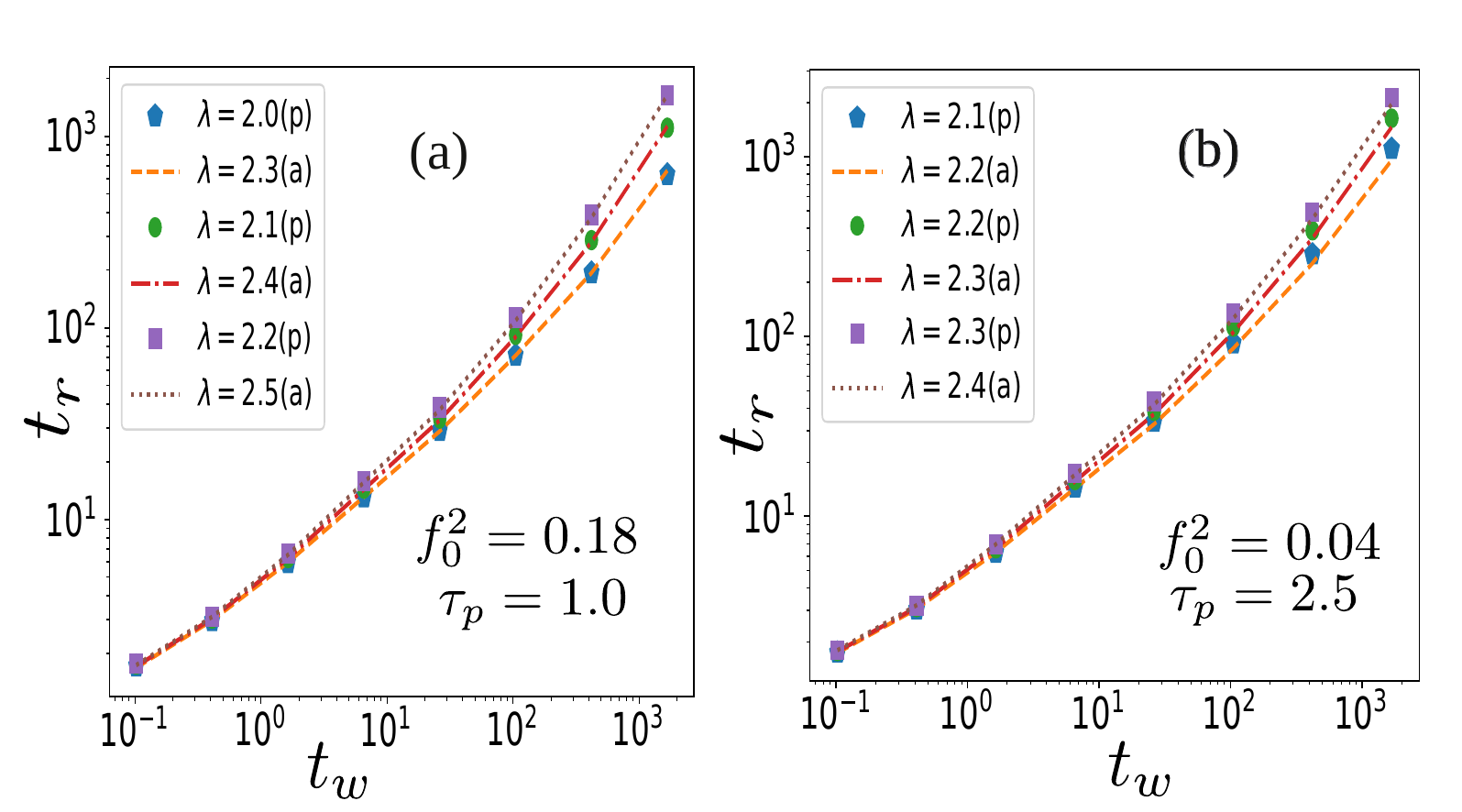}
	\caption{Aging is governed by distance from the respective critical points of the active and passive systems. We show this for two additional values of $\delta \lambda_c$: $\delta \lambda_c = 0.3$ (a) and $\delta \lambda_c = 0.1$ (b).}
	\label{supplement}
\end{figure}

Here we show $t_r$ vs $t_w$ plot demonstrating that distance from criticality governs the aging (Fig. \ref{supplement}).

\section{Analysis of existing simulation data for the exponent $\delta$}

\begin{figure}
	\centering
	\includegraphics[width=15cm,height=6.cm]{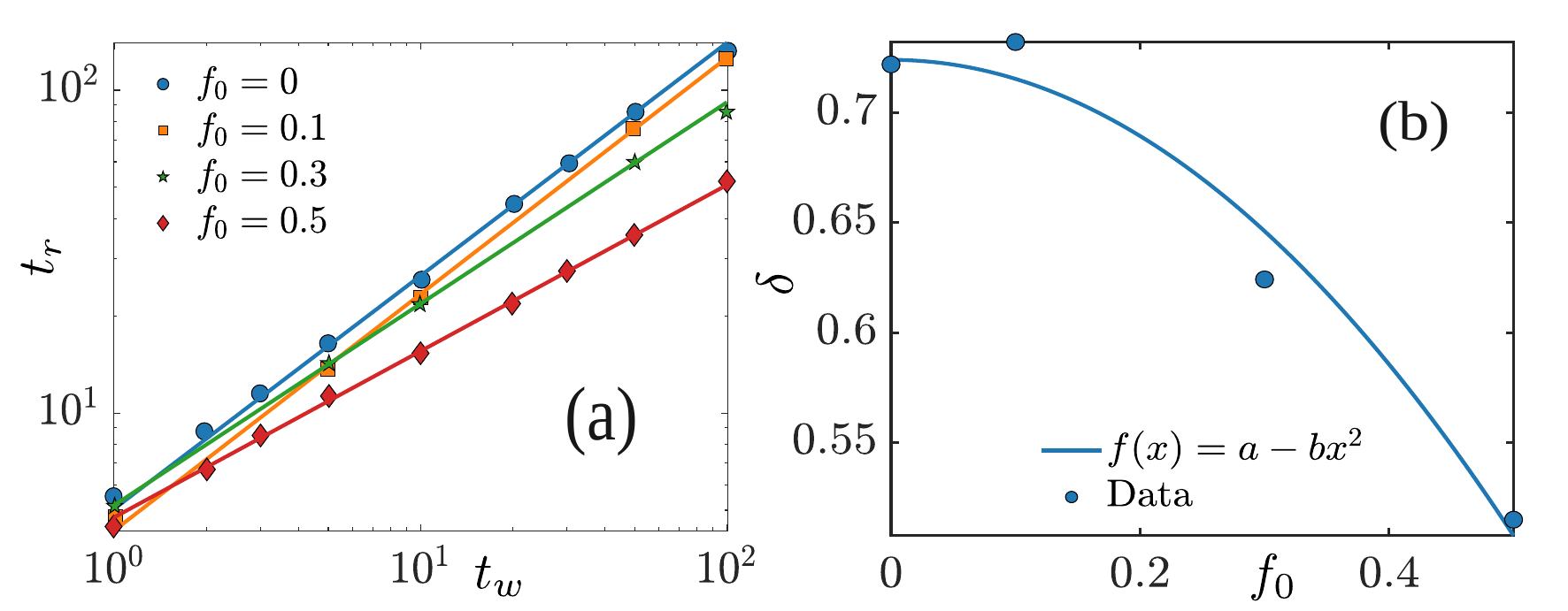}
	\caption{(a) Power-law fit of data from the study by Janzen and Janssen.  
		(b) The fitted exponent \( \delta \) exhibits a parabolic decay with increasing active force \( f_0 \). Line is the fit with the function $f(x) = a - bx^2$, where $a = 0.72$ and $b = 0.8$ and points are the exponent value from fig. (a).}
	\label{janzendata}
\end{figure}
We analyzed the data of Janzen and Janssen presented in the supplementary material, Fig. S5, in Ref. \cite{janzen2022aging} to obtain the exponent $\delta$. We collected the data of $t_r$ as a function of $t_w$ for various $f_0$ and fit the data with the power-law form: $t_r\sim t_w^\delta$. Figure \ref{janzendata}(a) shows the data from the paper by symbols and the fits by the lines. 
We plot the values of $\delta$ as a function of $f_0$ (Fig. \ref{janzendata}b) and fit with the function $f(x)=a-bx^2$ (line). The data is consistent with the prediction of the theory.

\section{Argument for the activity-dependence of the aging exponent, $\delta$}

We can write the characteristic time scale in terms of the barrier energy $E$ via the Arrhenius forms as
\begin{equation}
	\tau = \tau_0 \exp(\beta E),
\end{equation}
where $\tau_0$ is the high temperature microscopic time scale and $\beta$ is the inverse effective temperature (we have used the Boltzmann constant $k_B=1$).

In the aging regime, the relaxation time $\tau$ depends on the waiting time $t_w$. Therefore, $E$ will depend on $t_w$. We write the relaxation time in the passive (reference) case as:
\begin{equation}
	\tau = \tau_0 \exp(\beta E_0),
\end{equation}
and in the active case as:
\begin{equation}
	\tau^A = \tau_0 \exp(\beta E).
\end{equation}

Note that here $E < E_0$ as activity reduces the energy barrier \cite{nandi2018random}. 

Taking the ratio of the two expressions, we obtain:
\begin{equation}
	\frac{\tau^A}{\tau} = \exp\big[\beta (E - E_0)\big].
\end{equation}

Assuming a power-law dependence of the time scale on waiting time,
\begin{equation}
	\tau \sim t_w^{\delta_0}, \quad \tau^A \sim t_w^{\delta},
\end{equation}
we get:
\begin{equation}
	t_w^{(\delta - \delta_0)} = \exp\big[\beta (E - E_0)\big].
\end{equation}

Taking the logarithm, we obtain
\begin{equation}
	(\delta - \delta_0)\,\log t_w = \beta (E - E_0).
\end{equation}

Now we decompose the energy as:
\begin{equation}
	E = E_0 - E_A,
\end{equation}
where $E_A$ represents the activity contribution. Within the barrier crossing scenario, an energy scale should depend logarithmically on a time-scale, thus, $E_A=E_A(\log t_w)$.

Expanding $E_A(\log t_w)$ to leading order:
\begin{equation}
	E_A = E_A^0 + \Delta E_A \log t_w.
\end{equation}

Substituting back, we obtain:
\begin{equation}
	(\delta - \delta_0)\,\log t_w = - \beta \left(E_A^0 + \Delta E_A \log t_w \right).
\end{equation}

Comparing the coefficients of $\log t_w$, we finally get:
\begin{equation}
	\delta = \delta_0 - \beta \Delta E_A,
\end{equation}
where $\Delta E_A$ is proportional to an energy scale of the active system. We take it as the potential energy. Thus, $\Delta E\propto f_0^2\tau_p/(1+G\tau_p)$ for the ABP systems and  $\Delta E\propto f_0^2/(1+G\tau_p)$ for the AOUP systems \cite{nandi2017nonequilibrium, nandi2018random}. Thus, we obtain for ABP system,
\begin{equation} 
	\delta = \delta_0 - bf_0^2\tau_p/(1+G\tau_p),
\end{equation}
and for the AOUP system
\begin{equation} 
	\delta = \delta_0 - bf_0^2/(1+G\tau_p),
\end{equation}
where $b$ and $G$ are model dependent constants and $\delta_0$ is a $\lambda$ dependent constant \cite{warren2013quench}.


\begin{thebibliography}{82}%
\makeatletter
\providecommand \@ifxundefined [1]{%
 \@ifx{#1\undefined}
}%
\providecommand \@ifnum [1]{%
 \ifnum #1\expandafter \@firstoftwo
 \else \expandafter \@secondoftwo
 \fi
}%
\providecommand \@ifx [1]{%
 \ifx #1\expandafter \@firstoftwo
 \else \expandafter \@secondoftwo
 \fi
}%
\providecommand \natexlab [1]{#1}%
\providecommand \enquote  [1]{``#1''}%
\providecommand \bibnamefont  [1]{#1}%
\providecommand \bibfnamefont [1]{#1}%
\providecommand \citenamefont [1]{#1}%
\providecommand \href@noop [0]{\@secondoftwo}%
\providecommand \href [0]{\begingroup \@sanitize@url \@href}%
\providecommand \@href[1]{\@@startlink{#1}\@@href}%
\providecommand \@@href[1]{\endgroup#1\@@endlink}%
\providecommand \@sanitize@url [0]{\catcode `\\12\catcode `\$12\catcode
  `\&12\catcode `\#12\catcode `\^12\catcode `\_12\catcode `\%12\relax}%
\providecommand \@@startlink[1]{}%
\providecommand \@@endlink[0]{}%
\providecommand \url  [0]{\begingroup\@sanitize@url \@url }%
\providecommand \@url [1]{\endgroup\@href {#1}{\urlprefix }}%
\providecommand \urlprefix  [0]{URL }%
\providecommand \Eprint [0]{\href }%
\providecommand \doibase [0]{http://dx.doi.org/}%
\providecommand \selectlanguage [0]{\@gobble}%
\providecommand \bibinfo  [0]{\@secondoftwo}%
\providecommand \bibfield  [0]{\@secondoftwo}%
\providecommand \translation [1]{[#1]}%
\providecommand \BibitemOpen [0]{}%
\providecommand \bibitemStop [0]{}%
\providecommand \bibitemNoStop [0]{.\EOS\space}%
\providecommand \EOS [0]{\spacefactor3000\relax}%
\providecommand \BibitemShut  [1]{\csname bibitem#1\endcsname}%
\let\auto@bib@innerbib\@empty
\bibitem [{\citenamefont {Cugliandolo}\ and\ \citenamefont
  {Kurchan}(1993)}]{cugliandolo1993analytical}%
  \BibitemOpen
  \bibfield  {author} {\bibinfo {author} {\bibfnamefont {L.~F.}\ \bibnamefont
  {Cugliandolo}}\ and\ \bibinfo {author} {\bibfnamefont {J.}~\bibnamefont
  {Kurchan}},\ }\href {\doibase 10.1103/PhysRevLett.71.173} {\bibfield
  {journal} {\bibinfo  {journal} {Phys. Rev. Lett.}\ }\textbf {\bibinfo
  {volume} {71}},\ \bibinfo {pages} {173} (\bibinfo {year} {1993})}\BibitemShut
  {NoStop}%
\bibitem [{\citenamefont {Nandi}\ and\ \citenamefont
  {Ramaswamy}(2012)}]{nandi2012glassy}%
  \BibitemOpen
  \bibfield  {author} {\bibinfo {author} {\bibfnamefont {S.~K.}\ \bibnamefont
  {Nandi}}\ and\ \bibinfo {author} {\bibfnamefont {S.}~\bibnamefont
  {Ramaswamy}},\ }\href {\doibase PhysRevLett.109.115702} {\bibfield  {journal}
  {\bibinfo  {journal} {Phys. Rev. Lett.}\ }\textbf {\bibinfo {volume} {109}},\
  \bibinfo {pages} {115702} (\bibinfo {year} {2012})}\BibitemShut {NoStop}%
\bibitem [{\citenamefont {Nandi}\ and\ \citenamefont
  {Ramaswamy}(2016)}]{nandi2016glass}%
  \BibitemOpen
  \bibfield  {author} {\bibinfo {author} {\bibfnamefont {S.~K.}\ \bibnamefont
  {Nandi}}\ and\ \bibinfo {author} {\bibfnamefont {S.}~\bibnamefont
  {Ramaswamy}},\ }\href {\doibase 10.1103/PhysRevE.94.012607} {\bibfield
  {journal} {\bibinfo  {journal} {Phys. Rev. E}\ }\textbf {\bibinfo {volume}
  {94}},\ \bibinfo {pages} {012607} (\bibinfo {year} {2016})}\BibitemShut
  {NoStop}%
\bibitem [{\citenamefont {Lunkenheimer}\ \emph {et~al.}(2005)\citenamefont
  {Lunkenheimer}, \citenamefont {Wehn}, \citenamefont {Schneider},\ and\
  \citenamefont {Loidl}}]{lunkenheimer2005glassy}%
  \BibitemOpen
  \bibfield  {author} {\bibinfo {author} {\bibfnamefont {P.}~\bibnamefont
  {Lunkenheimer}}, \bibinfo {author} {\bibfnamefont {R.}~\bibnamefont {Wehn}},
  \bibinfo {author} {\bibfnamefont {U.}~\bibnamefont {Schneider}}, \ and\
  \bibinfo {author} {\bibfnamefont {A.}~\bibnamefont {Loidl}},\ }\href
  {\doibase 10.1103/PhysRevLett.95.055702} {\bibfield  {journal} {\bibinfo
  {journal} {Phys. Rev. Lett.}\ }\textbf {\bibinfo {volume} {95}},\ \bibinfo
  {pages} {055702} (\bibinfo {year} {2005})}\BibitemShut {NoStop}%
\bibitem [{\citenamefont {Abou}\ \emph {et~al.}(2001)\citenamefont {Abou},
  \citenamefont {Bonn},\ and\ \citenamefont {Meunier}}]{abou2001aging}%
  \BibitemOpen
  \bibfield  {author} {\bibinfo {author} {\bibfnamefont {B.}~\bibnamefont
  {Abou}}, \bibinfo {author} {\bibfnamefont {D.}~\bibnamefont {Bonn}}, \ and\
  \bibinfo {author} {\bibfnamefont {J.}~\bibnamefont {Meunier}},\ }\href
  {\doibase 10.1103/PhysRevE.64.021510} {\bibfield  {journal} {\bibinfo
  {journal} {Phys. Rev. E}\ }\textbf {\bibinfo {volume} {64}},\ \bibinfo
  {pages} {021510} (\bibinfo {year} {2001})}\BibitemShut {NoStop}%
\bibitem [{\citenamefont {Hodge}(1995)}]{hodge1995}%
  \BibitemOpen
  \bibfield  {author} {\bibinfo {author} {\bibfnamefont {I.~M.}\ \bibnamefont
  {Hodge}},\ }\href {\doibase 10.1126/science.267.5206.1945} {\bibfield
  {journal} {\bibinfo  {journal} {Science}\ }\textbf {\bibinfo {volume}
  {267}},\ \bibinfo {pages} {1945} (\bibinfo {year} {1995})}\BibitemShut
  {NoStop}%
\bibitem [{\citenamefont {Fabry}\ \emph {et~al.}(2001)\citenamefont {Fabry},
  \citenamefont {Maksym}, \citenamefont {Butler}, \citenamefont {Glogauer},
  \citenamefont {Navajas},\ and\ \citenamefont {Fredberg}}]{fabry2001scaling}%
  \BibitemOpen
  \bibfield  {author} {\bibinfo {author} {\bibfnamefont {B.}~\bibnamefont
  {Fabry}}, \bibinfo {author} {\bibfnamefont {G.~N.}\ \bibnamefont {Maksym}},
  \bibinfo {author} {\bibfnamefont {J.~P.}\ \bibnamefont {Butler}}, \bibinfo
  {author} {\bibfnamefont {M.}~\bibnamefont {Glogauer}}, \bibinfo {author}
  {\bibfnamefont {D.}~\bibnamefont {Navajas}}, \ and\ \bibinfo {author}
  {\bibfnamefont {J.~J.}\ \bibnamefont {Fredberg}},\ }\href {\doibase
  10.1103/PhysRevLett.87.148102} {\bibfield  {journal} {\bibinfo  {journal}
  {Phys. Rev. Lett.}\ }\textbf {\bibinfo {volume} {87}},\ \bibinfo {pages}
  {148102} (\bibinfo {year} {2001})}\BibitemShut {NoStop}%
\bibitem [{\citenamefont {Zhou}\ \emph {et~al.}(2009)\citenamefont {Zhou},
  \citenamefont {Trepat}, \citenamefont {Park}, \citenamefont {Lenormand},
  \citenamefont {Oliver}, \citenamefont {Mijailovich}, \citenamefont {Hardin},
  \citenamefont {Weitz}, \citenamefont {Butler},\ and\ \citenamefont
  {Fredberg}}]{zhou2009universal}%
  \BibitemOpen
  \bibfield  {author} {\bibinfo {author} {\bibfnamefont {E.}~\bibnamefont
  {Zhou}}, \bibinfo {author} {\bibfnamefont {X.}~\bibnamefont {Trepat}},
  \bibinfo {author} {\bibfnamefont {C.}~\bibnamefont {Park}}, \bibinfo {author}
  {\bibfnamefont {G.}~\bibnamefont {Lenormand}}, \bibinfo {author}
  {\bibfnamefont {M.}~\bibnamefont {Oliver}}, \bibinfo {author} {\bibfnamefont
  {S.}~\bibnamefont {Mijailovich}}, \bibinfo {author} {\bibfnamefont
  {C.}~\bibnamefont {Hardin}}, \bibinfo {author} {\bibfnamefont
  {D.}~\bibnamefont {Weitz}}, \bibinfo {author} {\bibfnamefont
  {J.}~\bibnamefont {Butler}}, \ and\ \bibinfo {author} {\bibfnamefont
  {J.}~\bibnamefont {Fredberg}},\ }\href {\doibase 10.1073/pnas.0901462106}
  {\bibfield  {journal} {\bibinfo  {journal} {Proc. Nat. Acad. Sci. (USA)}\
  }\textbf {\bibinfo {volume} {106}},\ \bibinfo {pages} {10632} (\bibinfo
  {year} {2009})}\BibitemShut {NoStop}%
\bibitem [{\citenamefont {Garcia}\ \emph {et~al.}(2015)\citenamefont {Garcia},
  \citenamefont {Hannezo}, \citenamefont {Elgeti}, \citenamefont {Joanny},
  \citenamefont {Silberzan},\ and\ \citenamefont {Gov}}]{garcia2015physics}%
  \BibitemOpen
  \bibfield  {author} {\bibinfo {author} {\bibfnamefont {S.}~\bibnamefont
  {Garcia}}, \bibinfo {author} {\bibfnamefont {E.}~\bibnamefont {Hannezo}},
  \bibinfo {author} {\bibfnamefont {J.}~\bibnamefont {Elgeti}}, \bibinfo
  {author} {\bibfnamefont {J.-F.}\ \bibnamefont {Joanny}}, \bibinfo {author}
  {\bibfnamefont {P.}~\bibnamefont {Silberzan}}, \ and\ \bibinfo {author}
  {\bibfnamefont {N.~S.}\ \bibnamefont {Gov}},\ }\href {\doibase
  10.1073/pnas.1510973112} {\bibfield  {journal} {\bibinfo  {journal} {Proc.
  Nat. Acad. Sci. (USA)}\ }\textbf {\bibinfo {volume} {112}},\ \bibinfo {pages}
  {15314} (\bibinfo {year} {2015})}\BibitemShut {NoStop}%
\bibitem [{\citenamefont {Angelini}\ \emph {et~al.}(2011)\citenamefont
  {Angelini}, \citenamefont {Hannezo}, \citenamefont {Trepat}, \citenamefont
  {Marquez}, \citenamefont {Fredberg},\ and\ \citenamefont
  {Weitz}}]{angelini2011glass}%
  \BibitemOpen
  \bibfield  {author} {\bibinfo {author} {\bibfnamefont {T.~E.}\ \bibnamefont
  {Angelini}}, \bibinfo {author} {\bibfnamefont {E.}~\bibnamefont {Hannezo}},
  \bibinfo {author} {\bibfnamefont {X.}~\bibnamefont {Trepat}}, \bibinfo
  {author} {\bibfnamefont {M.}~\bibnamefont {Marquez}}, \bibinfo {author}
  {\bibfnamefont {J.~J.}\ \bibnamefont {Fredberg}}, \ and\ \bibinfo {author}
  {\bibfnamefont {D.~A.}\ \bibnamefont {Weitz}},\ }\href {\doibase
  10.1073/pnas.1010059108} {\bibfield  {journal} {\bibinfo  {journal} {Proc.
  Nat. Acad. Sci. (USA)}\ }\textbf {\bibinfo {volume} {108}},\ \bibinfo {pages}
  {4714} (\bibinfo {year} {2011})}\BibitemShut {NoStop}%
\bibitem [{\citenamefont {Park}\ \emph {et~al.}(2015)\citenamefont {Park},
  \citenamefont {Kim}, \citenamefont {Bi}, \citenamefont {Mitchel},
  \citenamefont {Qazvini}, \citenamefont {Tantisira}, \citenamefont {Park},
  \citenamefont {McGill}, \citenamefont {Kim}, \citenamefont {Gweon} \emph
  {et~al.}}]{park2015unjamming}%
  \BibitemOpen
  \bibfield  {author} {\bibinfo {author} {\bibfnamefont {J.-A.}\ \bibnamefont
  {Park}}, \bibinfo {author} {\bibfnamefont {J.~H.}\ \bibnamefont {Kim}},
  \bibinfo {author} {\bibfnamefont {D.}~\bibnamefont {Bi}}, \bibinfo {author}
  {\bibfnamefont {J.~A.}\ \bibnamefont {Mitchel}}, \bibinfo {author}
  {\bibfnamefont {N.~T.}\ \bibnamefont {Qazvini}}, \bibinfo {author}
  {\bibfnamefont {K.}~\bibnamefont {Tantisira}}, \bibinfo {author}
  {\bibfnamefont {C.~Y.}\ \bibnamefont {Park}}, \bibinfo {author}
  {\bibfnamefont {M.}~\bibnamefont {McGill}}, \bibinfo {author} {\bibfnamefont
  {S.-H.}\ \bibnamefont {Kim}}, \bibinfo {author} {\bibfnamefont
  {B.}~\bibnamefont {Gweon}},  \emph {et~al.},\ }\href {\doibase
  10.1038/nmat4357} {\bibfield  {journal} {\bibinfo  {journal} {Nat. Mat.}\
  }\textbf {\bibinfo {volume} {14}},\ \bibinfo {pages} {1040} (\bibinfo {year}
  {2015})}\BibitemShut {NoStop}%
\bibitem [{\citenamefont {Malinverno}\ \emph {et~al.}(2017)\citenamefont
  {Malinverno}, \citenamefont {Corallino}, \citenamefont {Giavazzi},
  \citenamefont {Bergert}, \citenamefont {Li}, \citenamefont {Leoni},
  \citenamefont {Disanza}, \citenamefont {Frittoli}, \citenamefont {Oldani},
  \citenamefont {Martini} \emph {et~al.}}]{malinverno2017endocytic}%
  \BibitemOpen
  \bibfield  {author} {\bibinfo {author} {\bibfnamefont {C.}~\bibnamefont
  {Malinverno}}, \bibinfo {author} {\bibfnamefont {S.}~\bibnamefont
  {Corallino}}, \bibinfo {author} {\bibfnamefont {F.}~\bibnamefont {Giavazzi}},
  \bibinfo {author} {\bibfnamefont {M.}~\bibnamefont {Bergert}}, \bibinfo
  {author} {\bibfnamefont {Q.}~\bibnamefont {Li}}, \bibinfo {author}
  {\bibfnamefont {M.}~\bibnamefont {Leoni}}, \bibinfo {author} {\bibfnamefont
  {A.}~\bibnamefont {Disanza}}, \bibinfo {author} {\bibfnamefont
  {E.}~\bibnamefont {Frittoli}}, \bibinfo {author} {\bibfnamefont
  {A.}~\bibnamefont {Oldani}}, \bibinfo {author} {\bibfnamefont
  {E.}~\bibnamefont {Martini}},  \emph {et~al.},\ }\href {\doibase
  10.1038/nmat4848} {\bibfield  {journal} {\bibinfo  {journal} {Nat. Mat.}\
  }\textbf {\bibinfo {volume} {16}},\ \bibinfo {pages} {587} (\bibinfo {year}
  {2017})}\BibitemShut {NoStop}%
\bibitem [{\citenamefont {Atia}\ \emph {et~al.}(2018)\citenamefont {Atia},
  \citenamefont {Bi}, \citenamefont {Sharma}, \citenamefont {Mitchel},
  \citenamefont {Gweon}, \citenamefont {A.~Koehler}, \citenamefont {DeCamp},
  \citenamefont {Lan}, \citenamefont {Kim}, \citenamefont {Hirsch} \emph
  {et~al.}}]{atia2018geometric}%
  \BibitemOpen
  \bibfield  {author} {\bibinfo {author} {\bibfnamefont {L.}~\bibnamefont
  {Atia}}, \bibinfo {author} {\bibfnamefont {D.}~\bibnamefont {Bi}}, \bibinfo
  {author} {\bibfnamefont {Y.}~\bibnamefont {Sharma}}, \bibinfo {author}
  {\bibfnamefont {J.~A.}\ \bibnamefont {Mitchel}}, \bibinfo {author}
  {\bibfnamefont {B.}~\bibnamefont {Gweon}}, \bibinfo {author} {\bibfnamefont
  {S.}~\bibnamefont {A.~Koehler}}, \bibinfo {author} {\bibfnamefont {S.~J.}\
  \bibnamefont {DeCamp}}, \bibinfo {author} {\bibfnamefont {B.}~\bibnamefont
  {Lan}}, \bibinfo {author} {\bibfnamefont {J.~H.}\ \bibnamefont {Kim}},
  \bibinfo {author} {\bibfnamefont {R.}~\bibnamefont {Hirsch}},  \emph
  {et~al.},\ }\href {\doibase 10.1038/s41567-018-0089-9} {\bibfield  {journal}
  {\bibinfo  {journal} {Nat. phys.}\ }\textbf {\bibinfo {volume} {14}},\
  \bibinfo {pages} {613} (\bibinfo {year} {2018})}\BibitemShut {NoStop}%
\bibitem [{\citenamefont {Bursac}\ \emph {et~al.}(2005)\citenamefont {Bursac},
  \citenamefont {Lenormand}, \citenamefont {Fabry}, \citenamefont {Oliver},
  \citenamefont {Weitz}, \citenamefont {Viasnoff}, \citenamefont {Butler},\
  and\ \citenamefont {Fredberg}}]{bursac2005cytoskeletal}%
  \BibitemOpen
  \bibfield  {author} {\bibinfo {author} {\bibfnamefont {P.}~\bibnamefont
  {Bursac}}, \bibinfo {author} {\bibfnamefont {G.}~\bibnamefont {Lenormand}},
  \bibinfo {author} {\bibfnamefont {B.}~\bibnamefont {Fabry}}, \bibinfo
  {author} {\bibfnamefont {M.}~\bibnamefont {Oliver}}, \bibinfo {author}
  {\bibfnamefont {D.~A.}\ \bibnamefont {Weitz}}, \bibinfo {author}
  {\bibfnamefont {V.}~\bibnamefont {Viasnoff}}, \bibinfo {author}
  {\bibfnamefont {J.~P.}\ \bibnamefont {Butler}}, \ and\ \bibinfo {author}
  {\bibfnamefont {J.~J.}\ \bibnamefont {Fredberg}},\ }\href {\doibase
  10.1038/nmat1404} {\bibfield  {journal} {\bibinfo  {journal} {Nat. Mat.}\
  }\textbf {\bibinfo {volume} {4}},\ \bibinfo {pages} {557} (\bibinfo {year}
  {2005})}\BibitemShut {NoStop}%
\bibitem [{\citenamefont {Giavazzi}\ \emph {et~al.}(2018)\citenamefont
  {Giavazzi}, \citenamefont {Malinverno}, \citenamefont {Scita},\ and\
  \citenamefont {Cerbino}}]{giavazzi2018}%
  \BibitemOpen
  \bibfield  {author} {\bibinfo {author} {\bibfnamefont {F.}~\bibnamefont
  {Giavazzi}}, \bibinfo {author} {\bibfnamefont {C.}~\bibnamefont
  {Malinverno}}, \bibinfo {author} {\bibfnamefont {G.}~\bibnamefont {Scita}}, \
  and\ \bibinfo {author} {\bibfnamefont {R.}~\bibnamefont {Cerbino}},\ }\href
  {\doibase 10.3389/fphy.2018.00120} {\bibfield  {journal} {\bibinfo  {journal}
  {Front. Phys.}\ }\textbf {\bibinfo {volume} {6}},\ \bibinfo {pages} {120}
  (\bibinfo {year} {2018})}\BibitemShut {NoStop}%
\bibitem [{\citenamefont {Cerbino}\ \emph {et~al.}(2021)\citenamefont
  {Cerbino}, \citenamefont {Villa}, \citenamefont {Palamidessi}, \citenamefont
  {Frittoli}, \citenamefont {Scita},\ and\ \citenamefont
  {Giavazzi}}]{cerbino2021}%
  \BibitemOpen
  \bibfield  {author} {\bibinfo {author} {\bibfnamefont {R.}~\bibnamefont
  {Cerbino}}, \bibinfo {author} {\bibfnamefont {S.}~\bibnamefont {Villa}},
  \bibinfo {author} {\bibfnamefont {A.}~\bibnamefont {Palamidessi}}, \bibinfo
  {author} {\bibfnamefont {E.}~\bibnamefont {Frittoli}}, \bibinfo {author}
  {\bibfnamefont {G.}~\bibnamefont {Scita}}, \ and\ \bibinfo {author}
  {\bibfnamefont {F.}~\bibnamefont {Giavazzi}},\ }\href {\doibase
  10.1039/d0sm01837f} {\bibfield  {journal} {\bibinfo  {journal} {Soft Matter}\
  }\textbf {\bibinfo {volume} {17}},\ \bibinfo {pages} {3550} (\bibinfo {year}
  {2021})}\BibitemShut {NoStop}%
\bibitem [{\citenamefont {Nishizawa}\ \emph {et~al.}(2017)\citenamefont
  {Nishizawa}, \citenamefont {Fujiwara}, \citenamefont {Ikenaga}, \citenamefont
  {Nakajo}, \citenamefont {Yanagisawa},\ and\ \citenamefont
  {Mizuno}}]{nishizawa2017}%
  \BibitemOpen
  \bibfield  {author} {\bibinfo {author} {\bibfnamefont {K.}~\bibnamefont
  {Nishizawa}}, \bibinfo {author} {\bibfnamefont {K.}~\bibnamefont {Fujiwara}},
  \bibinfo {author} {\bibfnamefont {M.}~\bibnamefont {Ikenaga}}, \bibinfo
  {author} {\bibfnamefont {N.}~\bibnamefont {Nakajo}}, \bibinfo {author}
  {\bibfnamefont {M.}~\bibnamefont {Yanagisawa}}, \ and\ \bibinfo {author}
  {\bibfnamefont {D.}~\bibnamefont {Mizuno}},\ }\href {\doibase
  10.1038/s41598-017-14883-y} {\bibfield  {journal} {\bibinfo  {journal} {Sci.
  Rep.}\ }\textbf {\bibinfo {volume} {7}},\ \bibinfo {pages} {15143} (\bibinfo
  {year} {2017})}\BibitemShut {NoStop}%
\bibitem [{\citenamefont {Jawerth}\ \emph {et~al.}(2020)\citenamefont
  {Jawerth}, \citenamefont {Fischer-Friedrich}, \citenamefont {Saha},
  \citenamefont {Wang}, \citenamefont {Franzmann}, \citenamefont {Zhang},
  \citenamefont {Sachweh}, \citenamefont {Ruer}, \citenamefont {Ijavi},
  \citenamefont {Saha} \emph {et~al.}}]{jawerth2020protein}%
  \BibitemOpen
  \bibfield  {author} {\bibinfo {author} {\bibfnamefont {L.}~\bibnamefont
  {Jawerth}}, \bibinfo {author} {\bibfnamefont {E.}~\bibnamefont
  {Fischer-Friedrich}}, \bibinfo {author} {\bibfnamefont {S.}~\bibnamefont
  {Saha}}, \bibinfo {author} {\bibfnamefont {J.}~\bibnamefont {Wang}}, \bibinfo
  {author} {\bibfnamefont {T.}~\bibnamefont {Franzmann}}, \bibinfo {author}
  {\bibfnamefont {X.}~\bibnamefont {Zhang}}, \bibinfo {author} {\bibfnamefont
  {J.}~\bibnamefont {Sachweh}}, \bibinfo {author} {\bibfnamefont
  {M.}~\bibnamefont {Ruer}}, \bibinfo {author} {\bibfnamefont {M.}~\bibnamefont
  {Ijavi}}, \bibinfo {author} {\bibfnamefont {S.}~\bibnamefont {Saha}},  \emph
  {et~al.},\ }\href {\doibase 10.1126/science.aaw4951} {\bibfield  {journal}
  {\bibinfo  {journal} {Science}\ }\textbf {\bibinfo {volume} {370}},\ \bibinfo
  {pages} {1317} (\bibinfo {year} {2020})}\BibitemShut {NoStop}%
\bibitem [{\citenamefont {Alshareedah}\ \emph {et~al.}(2021)\citenamefont
  {Alshareedah}, \citenamefont {Moosa}, \citenamefont {Pham}, \citenamefont
  {Potoyan},\ and\ \citenamefont {Banerjee}}]{alshareedah2021programmable}%
  \BibitemOpen
  \bibfield  {author} {\bibinfo {author} {\bibfnamefont {I.}~\bibnamefont
  {Alshareedah}}, \bibinfo {author} {\bibfnamefont {M.~M.}\ \bibnamefont
  {Moosa}}, \bibinfo {author} {\bibfnamefont {M.}~\bibnamefont {Pham}},
  \bibinfo {author} {\bibfnamefont {D.~A.}\ \bibnamefont {Potoyan}}, \ and\
  \bibinfo {author} {\bibfnamefont {P.~R.}\ \bibnamefont {Banerjee}},\ }\href
  {\doibase 10.1038/s41467-021-26733-7} {\bibfield  {journal} {\bibinfo
  {journal} {Nat. Commun.}\ }\textbf {\bibinfo {volume} {12}},\ \bibinfo
  {pages} {6620} (\bibinfo {year} {2021})}\BibitemShut {NoStop}%
\bibitem [{\citenamefont {Takaki}\ \emph {et~al.}(2023)\citenamefont {Takaki},
  \citenamefont {Jawerth}, \citenamefont {Popovi{\'c}},\ and\ \citenamefont
  {J{\"u}licher}}]{takaki2023theory}%
  \BibitemOpen
  \bibfield  {author} {\bibinfo {author} {\bibfnamefont {R.}~\bibnamefont
  {Takaki}}, \bibinfo {author} {\bibfnamefont {L.}~\bibnamefont {Jawerth}},
  \bibinfo {author} {\bibfnamefont {M.}~\bibnamefont {Popovi{\'c}}}, \ and\
  \bibinfo {author} {\bibfnamefont {F.}~\bibnamefont {J{\"u}licher}},\ }\href
  {\doibase 10.1103/PRXLife.1.013006} {\bibfield  {journal} {\bibinfo
  {journal} {PRX Life}\ }\textbf {\bibinfo {volume} {1}},\ \bibinfo {pages}
  {013006} (\bibinfo {year} {2023})}\BibitemShut {NoStop}%
\bibitem [{\citenamefont {Deng}\ \emph {et~al.}(2006)\citenamefont {Deng},
  \citenamefont {Trepat}, \citenamefont {Butler}, \citenamefont {Millet},
  \citenamefont {Morgan}, \citenamefont {Weitz},\ and\ \citenamefont
  {Fredberg}}]{deng2006fast}%
  \BibitemOpen
  \bibfield  {author} {\bibinfo {author} {\bibfnamefont {L.}~\bibnamefont
  {Deng}}, \bibinfo {author} {\bibfnamefont {X.}~\bibnamefont {Trepat}},
  \bibinfo {author} {\bibfnamefont {J.~P.}\ \bibnamefont {Butler}}, \bibinfo
  {author} {\bibfnamefont {E.}~\bibnamefont {Millet}}, \bibinfo {author}
  {\bibfnamefont {K.~G.}\ \bibnamefont {Morgan}}, \bibinfo {author}
  {\bibfnamefont {D.~A.}\ \bibnamefont {Weitz}}, \ and\ \bibinfo {author}
  {\bibfnamefont {J.~J.}\ \bibnamefont {Fredberg}},\ }\href {\doibase
  10.1038/nmat1685} {\bibfield  {journal} {\bibinfo  {journal} {Nat. Mat.}\
  }\textbf {\bibinfo {volume} {5}},\ \bibinfo {pages} {636} (\bibinfo {year}
  {2006})}\BibitemShut {NoStop}%
\bibitem [{\citenamefont {Sadhukhan}\ and\ \citenamefont
  {Nandi}(2021)}]{sadhukhan2021theory}%
  \BibitemOpen
  \bibfield  {author} {\bibinfo {author} {\bibfnamefont {S.}~\bibnamefont
  {Sadhukhan}}\ and\ \bibinfo {author} {\bibfnamefont {S.~K.}\ \bibnamefont
  {Nandi}},\ }\href {\doibase 10.1103/PhysRevE.103.062403} {\bibfield
  {journal} {\bibinfo  {journal} {Phys. Rev. E}\ }\textbf {\bibinfo {volume}
  {103}},\ \bibinfo {pages} {062403} (\bibinfo {year} {2021})}\BibitemShut
  {NoStop}%
\bibitem [{\citenamefont {Takatori}\ and\ \citenamefont
  {Mandadapu}(2020)}]{takatori2020motility}%
  \BibitemOpen
  \bibfield  {author} {\bibinfo {author} {\bibfnamefont {S.~C.}\ \bibnamefont
  {Takatori}}\ and\ \bibinfo {author} {\bibfnamefont {K.~K.}\ \bibnamefont
  {Mandadapu}},\ }\href {\doibase https://doi.org/10.48550/arXiv.2003.05618}
  {\bibfield  {journal} {\bibinfo  {journal} {arXiv preprint arXiv:2003.05618}\
  } (\bibinfo {year} {2020}),\
  https://doi.org/10.48550/arXiv.2003.05618}\BibitemShut {NoStop}%
\bibitem [{\citenamefont {Lama}\ \emph {et~al.}(2024)\citenamefont {Lama},
  \citenamefont {Yamamoto}, \citenamefont {Furuta}, \citenamefont {Shimaya},\
  and\ \citenamefont {Takeuchi}}]{lama2024emergence}%
  \BibitemOpen
  \bibfield  {author} {\bibinfo {author} {\bibfnamefont {H.}~\bibnamefont
  {Lama}}, \bibinfo {author} {\bibfnamefont {M.~J.}\ \bibnamefont {Yamamoto}},
  \bibinfo {author} {\bibfnamefont {Y.}~\bibnamefont {Furuta}}, \bibinfo
  {author} {\bibfnamefont {T.}~\bibnamefont {Shimaya}}, \ and\ \bibinfo
  {author} {\bibfnamefont {K.~A.}\ \bibnamefont {Takeuchi}},\ }\href {\doibase
  10.1093/pnasnexus/pgae238} {\bibfield  {journal} {\bibinfo  {journal} {PNAS
  Nexus}\ ,\ \bibinfo {pages} {pgae238}} (\bibinfo {year} {2024})}\BibitemShut
  {NoStop}%
\bibitem [{\citenamefont {Klongvessa}\ \emph {et~al.}(2019)\citenamefont
  {Klongvessa}, \citenamefont {Ginot}, \citenamefont {Ybert}, \citenamefont
  {Cottin-Bizonne},\ and\ \citenamefont {Leocmach}}]{klongvessa2019}%
  \BibitemOpen
  \bibfield  {author} {\bibinfo {author} {\bibfnamefont {N.}~\bibnamefont
  {Klongvessa}}, \bibinfo {author} {\bibfnamefont {F.}~\bibnamefont {Ginot}},
  \bibinfo {author} {\bibfnamefont {C.}~\bibnamefont {Ybert}}, \bibinfo
  {author} {\bibfnamefont {C.}~\bibnamefont {Cottin-Bizonne}}, \ and\ \bibinfo
  {author} {\bibfnamefont {M.}~\bibnamefont {Leocmach}},\ }\href {\doibase
  10.1103/physrevlett.123.248004} {\bibfield  {journal} {\bibinfo  {journal}
  {Phys. Rev. Lett.}\ }\textbf {\bibinfo {volume} {123}},\ \bibinfo {pages}
  {248004} (\bibinfo {year} {2019})}\BibitemShut {NoStop}%
\bibitem [{\citenamefont {Klongvessa}\ \emph
  {et~al.}(2022{\natexlab{a}})\citenamefont {Klongvessa}, \citenamefont
  {Ybert}, \citenamefont {Cottin-Bizonne}, \citenamefont {Kawasaki},\ and\
  \citenamefont {Leocmach}}]{klongvessa2022}%
  \BibitemOpen
  \bibfield  {author} {\bibinfo {author} {\bibfnamefont {N.}~\bibnamefont
  {Klongvessa}}, \bibinfo {author} {\bibfnamefont {C.}~\bibnamefont {Ybert}},
  \bibinfo {author} {\bibfnamefont {C.}~\bibnamefont {Cottin-Bizonne}},
  \bibinfo {author} {\bibfnamefont {T.}~\bibnamefont {Kawasaki}}, \ and\
  \bibinfo {author} {\bibfnamefont {M.}~\bibnamefont {Leocmach}},\ }\href
  {\doibase https://doi.org/10.1063/5.0087578} {\bibfield  {journal} {\bibinfo
  {journal} {J. Chem. Phys.}\ }\textbf {\bibinfo {volume} {156}},\ \bibinfo
  {pages} {154509} (\bibinfo {year} {2022}{\natexlab{a}})}\BibitemShut
  {NoStop}%
\bibitem [{\citenamefont {Arora}\ \emph {et~al.}(2022)\citenamefont {Arora},
  \citenamefont {Sood},\ and\ \citenamefont {Ganapathy}}]{arora2022}%
  \BibitemOpen
  \bibfield  {author} {\bibinfo {author} {\bibfnamefont {P.}~\bibnamefont
  {Arora}}, \bibinfo {author} {\bibfnamefont {A.~K.}\ \bibnamefont {Sood}}, \
  and\ \bibinfo {author} {\bibfnamefont {R.}~\bibnamefont {Ganapathy}},\ }\href
  {\doibase 10.1103/PhysRevLett.128.178002} {\bibfield  {journal} {\bibinfo
  {journal} {Phys. Rev. Lett.}\ }\textbf {\bibinfo {volume} {128}},\ \bibinfo
  {pages} {178002} (\bibinfo {year} {2022})}\BibitemShut {NoStop}%
\bibitem [{\citenamefont {Ghosh}\ \emph {et~al.}(2024)\citenamefont {Ghosh},
  \citenamefont {Maity},\ and\ \citenamefont {Chikkadi}}]{vijay2024}%
  \BibitemOpen
  \bibfield  {author} {\bibinfo {author} {\bibfnamefont {A.}~\bibnamefont
  {Ghosh}}, \bibinfo {author} {\bibfnamefont {S.}~\bibnamefont {Maity}}, \ and\
  \bibinfo {author} {\bibfnamefont {V.}~\bibnamefont {Chikkadi}},\ }\href
  {\doibase 10.48550/arXiv.2406.17927} {\bibfield  {journal} {\bibinfo
  {journal} {arXiv}\ ,\ \bibinfo {pages} {2406.17927}} (\bibinfo {year}
  {2024})}\BibitemShut {NoStop}%
\bibitem [{\citenamefont {Poujade}\ \emph {et~al.}(2007)\citenamefont
  {Poujade}, \citenamefont {Grasland-Mongrain}, \citenamefont {Hertzog},
  \citenamefont {Jouanneau}, \citenamefont {Chavrier}, \citenamefont {Ladoux},
  \citenamefont {Buguin},\ and\ \citenamefont
  {Silberzan}}]{poujade2007collective}%
  \BibitemOpen
  \bibfield  {author} {\bibinfo {author} {\bibfnamefont {M.}~\bibnamefont
  {Poujade}}, \bibinfo {author} {\bibfnamefont {E.}~\bibnamefont
  {Grasland-Mongrain}}, \bibinfo {author} {\bibfnamefont {A.}~\bibnamefont
  {Hertzog}}, \bibinfo {author} {\bibfnamefont {J.}~\bibnamefont {Jouanneau}},
  \bibinfo {author} {\bibfnamefont {P.}~\bibnamefont {Chavrier}}, \bibinfo
  {author} {\bibfnamefont {B.}~\bibnamefont {Ladoux}}, \bibinfo {author}
  {\bibfnamefont {A.}~\bibnamefont {Buguin}}, \ and\ \bibinfo {author}
  {\bibfnamefont {P.}~\bibnamefont {Silberzan}},\ }\href {\doibase
  10.1073/pnas.0705062104} {\bibfield  {journal} {\bibinfo  {journal} {Proc.
  Nat. Acad. Sci. (USA)}\ }\textbf {\bibinfo {volume} {104}},\ \bibinfo {pages}
  {15988} (\bibinfo {year} {2007})}\BibitemShut {NoStop}%
\bibitem [{\citenamefont {Brugu{\'e}s}\ \emph {et~al.}(2014)\citenamefont
  {Brugu{\'e}s}, \citenamefont {Anon}, \citenamefont {Conte}, \citenamefont
  {Veldhuis}, \citenamefont {Gupta}, \citenamefont {Colombelli}, \citenamefont
  {Mu{\~n}oz}, \citenamefont {Brodland}, \citenamefont {Ladoux},\ and\
  \citenamefont {Trepat}}]{brugues2014forces}%
  \BibitemOpen
  \bibfield  {author} {\bibinfo {author} {\bibfnamefont {A.}~\bibnamefont
  {Brugu{\'e}s}}, \bibinfo {author} {\bibfnamefont {E.}~\bibnamefont {Anon}},
  \bibinfo {author} {\bibfnamefont {V.}~\bibnamefont {Conte}}, \bibinfo
  {author} {\bibfnamefont {J.~H.}\ \bibnamefont {Veldhuis}}, \bibinfo {author}
  {\bibfnamefont {M.}~\bibnamefont {Gupta}}, \bibinfo {author} {\bibfnamefont
  {J.}~\bibnamefont {Colombelli}}, \bibinfo {author} {\bibfnamefont {J.~J.}\
  \bibnamefont {Mu{\~n}oz}}, \bibinfo {author} {\bibfnamefont {G.~W.}\
  \bibnamefont {Brodland}}, \bibinfo {author} {\bibfnamefont {B.}~\bibnamefont
  {Ladoux}}, \ and\ \bibinfo {author} {\bibfnamefont {X.}~\bibnamefont
  {Trepat}},\ }\href {\doibase 10.1038/nphys3040} {\bibfield  {journal}
  {\bibinfo  {journal} {Nat. phys.}\ }\textbf {\bibinfo {volume} {10}},\
  \bibinfo {pages} {683} (\bibinfo {year} {2014})}\BibitemShut {NoStop}%
\bibitem [{\citenamefont {Friedl}\ and\ \citenamefont
  {Gilmour}(2009)}]{friedl2009collective}%
  \BibitemOpen
  \bibfield  {author} {\bibinfo {author} {\bibfnamefont {P.}~\bibnamefont
  {Friedl}}\ and\ \bibinfo {author} {\bibfnamefont {D.}~\bibnamefont
  {Gilmour}},\ }\href {\doibase 10.1038/nrm2720} {\bibfield  {journal}
  {\bibinfo  {journal} {Nat. Rev. Mol. Cell Biol.}\ }\textbf {\bibinfo {volume}
  {10}},\ \bibinfo {pages} {445} (\bibinfo {year} {2009})}\BibitemShut
  {NoStop}%
\bibitem [{\citenamefont {Tambe}\ \emph {et~al.}(2011)\citenamefont {Tambe},
  \citenamefont {Corey~Hardin}, \citenamefont {Angelini}, \citenamefont
  {Rajendran}, \citenamefont {Park}, \citenamefont {Serra-Picamal},
  \citenamefont {Zhou}, \citenamefont {Zaman}, \citenamefont {Butler},
  \citenamefont {Weitz} \emph {et~al.}}]{tambe2011collective}%
  \BibitemOpen
  \bibfield  {author} {\bibinfo {author} {\bibfnamefont {D.~T.}\ \bibnamefont
  {Tambe}}, \bibinfo {author} {\bibfnamefont {C.}~\bibnamefont {Corey~Hardin}},
  \bibinfo {author} {\bibfnamefont {T.~E.}\ \bibnamefont {Angelini}}, \bibinfo
  {author} {\bibfnamefont {K.}~\bibnamefont {Rajendran}}, \bibinfo {author}
  {\bibfnamefont {C.~Y.}\ \bibnamefont {Park}}, \bibinfo {author}
  {\bibfnamefont {X.}~\bibnamefont {Serra-Picamal}}, \bibinfo {author}
  {\bibfnamefont {E.~H.}\ \bibnamefont {Zhou}}, \bibinfo {author}
  {\bibfnamefont {M.~H.}\ \bibnamefont {Zaman}}, \bibinfo {author}
  {\bibfnamefont {J.~P.}\ \bibnamefont {Butler}}, \bibinfo {author}
  {\bibfnamefont {D.~A.}\ \bibnamefont {Weitz}},  \emph {et~al.},\ }\href
  {\doibase 10.1038/nmat3025} {\bibfield  {journal} {\bibinfo  {journal} {Nat.
  Mat.}\ }\textbf {\bibinfo {volume} {10}},\ \bibinfo {pages} {469} (\bibinfo
  {year} {2011})}\BibitemShut {NoStop}%
\bibitem [{\citenamefont {Malmi-Kakkada}\ \emph {et~al.}(2018)\citenamefont
  {Malmi-Kakkada}, \citenamefont {Li}, \citenamefont {Samanta}, \citenamefont
  {Sinha},\ and\ \citenamefont {Thirumalai}}]{malmi2018cell}%
  \BibitemOpen
  \bibfield  {author} {\bibinfo {author} {\bibfnamefont {A.~N.}\ \bibnamefont
  {Malmi-Kakkada}}, \bibinfo {author} {\bibfnamefont {X.}~\bibnamefont {Li}},
  \bibinfo {author} {\bibfnamefont {H.~S.}\ \bibnamefont {Samanta}}, \bibinfo
  {author} {\bibfnamefont {S.}~\bibnamefont {Sinha}}, \ and\ \bibinfo {author}
  {\bibfnamefont {D.}~\bibnamefont {Thirumalai}},\ }\href {\doibase
  10.1103/PhysRevX.8.021025} {\bibfield  {journal} {\bibinfo  {journal} {Phys.
  Rev. X}\ }\textbf {\bibinfo {volume} {8}},\ \bibinfo {pages} {021025}
  (\bibinfo {year} {2018})}\BibitemShut {NoStop}%
\bibitem [{\citenamefont {Sadhukhan}\ \emph
  {et~al.}(2024{\natexlab{a}})\citenamefont {Sadhukhan}, \citenamefont {Dey},
  \citenamefont {Karmakar},\ and\ \citenamefont
  {Nandi}}]{sadhukhan2024perspective}%
  \BibitemOpen
  \bibfield  {author} {\bibinfo {author} {\bibfnamefont {S.}~\bibnamefont
  {Sadhukhan}}, \bibinfo {author} {\bibfnamefont {S.}~\bibnamefont {Dey}},
  \bibinfo {author} {\bibfnamefont {S.}~\bibnamefont {Karmakar}}, \ and\
  \bibinfo {author} {\bibfnamefont {S.~K.}\ \bibnamefont {Nandi}},\ }\href
  {\doibase 10.1140/epjs/s11734-024-01188-1} {\bibfield  {journal} {\bibinfo
  {journal} {Euro. Phys. J. Spe. Top.}\ }\textbf {\bibinfo {volume} {233}},\
  \bibinfo {pages} {3193} (\bibinfo {year} {2024}{\natexlab{a}})}\BibitemShut
  {NoStop}%
\bibitem [{\citenamefont {Maiuri}\ \emph {et~al.}(2015)\citenamefont {Maiuri},
  \citenamefont {Rupprecht}, \citenamefont {Wieser}, \citenamefont {Ruprecht},
  \citenamefont {B{\'e}nichou}, \citenamefont {Carpi}, \citenamefont {Coppey},
  \citenamefont {De~Beco}, \citenamefont {Gov}, \citenamefont {Heisenberg}
  \emph {et~al.}}]{maiuri2015actin}%
  \BibitemOpen
  \bibfield  {author} {\bibinfo {author} {\bibfnamefont {P.}~\bibnamefont
  {Maiuri}}, \bibinfo {author} {\bibfnamefont {J.-F.}\ \bibnamefont
  {Rupprecht}}, \bibinfo {author} {\bibfnamefont {S.}~\bibnamefont {Wieser}},
  \bibinfo {author} {\bibfnamefont {V.}~\bibnamefont {Ruprecht}}, \bibinfo
  {author} {\bibfnamefont {O.}~\bibnamefont {B{\'e}nichou}}, \bibinfo {author}
  {\bibfnamefont {N.}~\bibnamefont {Carpi}}, \bibinfo {author} {\bibfnamefont
  {M.}~\bibnamefont {Coppey}}, \bibinfo {author} {\bibfnamefont
  {S.}~\bibnamefont {De~Beco}}, \bibinfo {author} {\bibfnamefont
  {N.}~\bibnamefont {Gov}}, \bibinfo {author} {\bibfnamefont {C.-P.}\
  \bibnamefont {Heisenberg}},  \emph {et~al.},\ }\href {\doibase
  10.1016/j.cell.2015.01.056} {\bibfield  {journal} {\bibinfo  {journal}
  {Cell}\ }\textbf {\bibinfo {volume} {161}},\ \bibinfo {pages} {374} (\bibinfo
  {year} {2015})}\BibitemShut {NoStop}%
\bibitem [{\citenamefont {Wortel}\ \emph {et~al.}(2021)\citenamefont {Wortel},
  \citenamefont {Niculescu}, \citenamefont {Kolijn}, \citenamefont {Gov},
  \citenamefont {de~Boer},\ and\ \citenamefont {Textor}}]{wortel2021}%
  \BibitemOpen
  \bibfield  {author} {\bibinfo {author} {\bibfnamefont {I.~M.}\ \bibnamefont
  {Wortel}}, \bibinfo {author} {\bibfnamefont {I.}~\bibnamefont {Niculescu}},
  \bibinfo {author} {\bibfnamefont {P.~M.}\ \bibnamefont {Kolijn}}, \bibinfo
  {author} {\bibfnamefont {N.~S.}\ \bibnamefont {Gov}}, \bibinfo {author}
  {\bibfnamefont {R.~J.}\ \bibnamefont {de~Boer}}, \ and\ \bibinfo {author}
  {\bibfnamefont {J.}~\bibnamefont {Textor}},\ }\href {\doibase
  10.1016/j.bpj.2021.04.036} {\bibfield  {journal} {\bibinfo  {journal}
  {Biophys. J.}\ }\textbf {\bibinfo {volume} {120}},\ \bibinfo {pages} {2609}
  (\bibinfo {year} {2021})}\BibitemShut {NoStop}%
\bibitem [{\citenamefont {Janssen}(2019)}]{janssen2019active}%
  \BibitemOpen
  \bibfield  {author} {\bibinfo {author} {\bibfnamefont {L.~M.}\ \bibnamefont
  {Janssen}},\ }\href {\doibase 10.1088/1361-648X/ab3e90} {\bibfield  {journal}
  {\bibinfo  {journal} {J. Phys.: Condens. Matter}\ }\textbf {\bibinfo {volume}
  {31}},\ \bibinfo {pages} {503002} (\bibinfo {year} {2019})}\BibitemShut
  {NoStop}%
\bibitem [{\citenamefont {Berthier}\ \emph {et~al.}(2019)\citenamefont
  {Berthier}, \citenamefont {Flenner},\ and\ \citenamefont
  {Szamel}}]{berthier2019glassy}%
  \BibitemOpen
  \bibfield  {author} {\bibinfo {author} {\bibfnamefont {L.}~\bibnamefont
  {Berthier}}, \bibinfo {author} {\bibfnamefont {E.}~\bibnamefont {Flenner}}, \
  and\ \bibinfo {author} {\bibfnamefont {G.}~\bibnamefont {Szamel}},\ }\href
  {\doibase 10.1063/1.5093240} {\bibfield  {journal} {\bibinfo  {journal} {J.
  Chem. Phys.}\ }\textbf {\bibinfo {volume} {150}},\ \bibinfo {pages} {200901}
  (\bibinfo {year} {2019})}\BibitemShut {NoStop}%
\bibitem [{\citenamefont {Berthier}\ and\ \citenamefont
  {Kurchan}(2013)}]{berthier2013}%
  \BibitemOpen
  \bibfield  {author} {\bibinfo {author} {\bibfnamefont {L.}~\bibnamefont
  {Berthier}}\ and\ \bibinfo {author} {\bibfnamefont {J.}~\bibnamefont
  {Kurchan}},\ }\href {\doibase 10.1038/nphys2592} {\bibfield  {journal}
  {\bibinfo  {journal} {Nat. Phys.}\ }\textbf {\bibinfo {volume} {9}},\
  \bibinfo {pages} {310} (\bibinfo {year} {2013})}\BibitemShut {NoStop}%
\bibitem [{\citenamefont {Szamel}\ \emph {et~al.}(2015)\citenamefont {Szamel},
  \citenamefont {Flenner},\ and\ \citenamefont {Berthier}}]{szamel2015}%
  \BibitemOpen
  \bibfield  {author} {\bibinfo {author} {\bibfnamefont {G.}~\bibnamefont
  {Szamel}}, \bibinfo {author} {\bibfnamefont {E.}~\bibnamefont {Flenner}}, \
  and\ \bibinfo {author} {\bibfnamefont {L.}~\bibnamefont {Berthier}},\ }\href
  {\doibase 10.1103/physreve.91.062304} {\bibfield  {journal} {\bibinfo
  {journal} {Phys. Rev. E}\ }\textbf {\bibinfo {volume} {91}},\ \bibinfo
  {pages} {062304} (\bibinfo {year} {2015})}\BibitemShut {NoStop}%
\bibitem [{\citenamefont {Liluashvili}\ \emph {et~al.}(2017)\citenamefont
  {Liluashvili}, \citenamefont {{\'{O}}nody},\ and\ \citenamefont
  {Voigtmann}}]{liluashvili2017}%
  \BibitemOpen
  \bibfield  {author} {\bibinfo {author} {\bibfnamefont {A.}~\bibnamefont
  {Liluashvili}}, \bibinfo {author} {\bibfnamefont {J.}~\bibnamefont
  {{\'{O}}nody}}, \ and\ \bibinfo {author} {\bibfnamefont {T.}~\bibnamefont
  {Voigtmann}},\ }\href {\doibase 10.1103/PhysRevE.96.062608} {\bibfield
  {journal} {\bibinfo  {journal} {Phys. Rev. E}\ }\textbf {\bibinfo {volume}
  {96}},\ \bibinfo {pages} {062608} (\bibinfo {year} {2017})}\BibitemShut
  {NoStop}%
\bibitem [{\citenamefont {Feng}\ and\ \citenamefont {Hou}(2017)}]{feng2017}%
  \BibitemOpen
  \bibfield  {author} {\bibinfo {author} {\bibfnamefont {M.}~\bibnamefont
  {Feng}}\ and\ \bibinfo {author} {\bibfnamefont {Z.}~\bibnamefont {Hou}},\
  }\href {\doibase 10.1039/C7SM00852J} {\bibfield  {journal} {\bibinfo
  {journal} {Soft Matter}\ }\textbf {\bibinfo {volume} {13}},\ \bibinfo {pages}
  {4464} (\bibinfo {year} {2017})}\BibitemShut {NoStop}%
\bibitem [{\citenamefont {Debets}\ and\ \citenamefont
  {Janssen}(2022)}]{debets2022}%
  \BibitemOpen
  \bibfield  {author} {\bibinfo {author} {\bibfnamefont {V.~E.}\ \bibnamefont
  {Debets}}\ and\ \bibinfo {author} {\bibfnamefont {L.~M.~C.}\ \bibnamefont
  {Janssen}},\ }\href {\doibase https://doi.org/10.1063/5.0127569} {\bibfield
  {journal} {\bibinfo  {journal} {J. Chem. Phys.}\ }\textbf {\bibinfo {volume}
  {157}},\ \bibinfo {pages} {224902} (\bibinfo {year} {2022})}\BibitemShut
  {NoStop}%
\bibitem [{\citenamefont {Nandi}\ and\ \citenamefont
  {Gov}(2017)}]{nandi2017nonequilibrium}%
  \BibitemOpen
  \bibfield  {author} {\bibinfo {author} {\bibfnamefont {S.~K.}\ \bibnamefont
  {Nandi}}\ and\ \bibinfo {author} {\bibfnamefont {N.~S.}\ \bibnamefont
  {Gov}},\ }\href {\doibase 10.1039/C7SM01648D} {\bibfield  {journal} {\bibinfo
   {journal} {Soft matter}\ }\textbf {\bibinfo {volume} {13}},\ \bibinfo
  {pages} {7609} (\bibinfo {year} {2017})}\BibitemShut {NoStop}%
\bibitem [{\citenamefont {Nandi}\ \emph {et~al.}(2018)\citenamefont {Nandi},
  \citenamefont {Mandal}, \citenamefont {Bhuyan}, \citenamefont {Dasgupta},
  \citenamefont {Rao},\ and\ \citenamefont {Gov}}]{nandi2018random}%
  \BibitemOpen
  \bibfield  {author} {\bibinfo {author} {\bibfnamefont {S.~K.}\ \bibnamefont
  {Nandi}}, \bibinfo {author} {\bibfnamefont {R.}~\bibnamefont {Mandal}},
  \bibinfo {author} {\bibfnamefont {P.~J.}\ \bibnamefont {Bhuyan}}, \bibinfo
  {author} {\bibfnamefont {C.}~\bibnamefont {Dasgupta}}, \bibinfo {author}
  {\bibfnamefont {M.}~\bibnamefont {Rao}}, \ and\ \bibinfo {author}
  {\bibfnamefont {N.~S.}\ \bibnamefont {Gov}},\ }\href {\doibase
  10.1073/pnas.1721324115} {\bibfield  {journal} {\bibinfo  {journal} {Proc.
  Nat. Acad. Sci. (USA)}\ }\textbf {\bibinfo {volume} {115}},\ \bibinfo {pages}
  {7688} (\bibinfo {year} {2018})}\BibitemShut {NoStop}%
\bibitem [{\citenamefont {Paul}\ \emph {et~al.}(2023)\citenamefont {Paul},
  \citenamefont {Mutneja}, \citenamefont {Nandi},\ and\ \citenamefont
  {Karmakar}}]{paul2023dynamical}%
  \BibitemOpen
  \bibfield  {author} {\bibinfo {author} {\bibfnamefont {K.}~\bibnamefont
  {Paul}}, \bibinfo {author} {\bibfnamefont {A.}~\bibnamefont {Mutneja}},
  \bibinfo {author} {\bibfnamefont {S.~K.}\ \bibnamefont {Nandi}}, \ and\
  \bibinfo {author} {\bibfnamefont {S.}~\bibnamefont {Karmakar}},\ }\href
  {\doibase 10.1073/pnas.2217073120} {\bibfield  {journal} {\bibinfo  {journal}
  {Proc. Nat. Acad. Sci. (USA}\ }\textbf {\bibinfo {volume} {120}},\ \bibinfo
  {pages} {e2217073120} (\bibinfo {year} {2023})}\BibitemShut {NoStop}%
\bibitem [{\citenamefont {Kolya}\ \emph {et~al.}(2024)\citenamefont {Kolya},
  \citenamefont {Pareek},\ and\ \citenamefont {Nandi}}]{kolya2024active}%
  \BibitemOpen
  \bibfield  {author} {\bibinfo {author} {\bibfnamefont {S.}~\bibnamefont
  {Kolya}}, \bibinfo {author} {\bibfnamefont {P.}~\bibnamefont {Pareek}}, \
  and\ \bibinfo {author} {\bibfnamefont {S.~K.}\ \bibnamefont {Nandi}},\
  }\href@noop {} {\bibfield  {journal} {\bibinfo  {journal} {arXiv preprint
  arXiv:2410.15928}\ } (\bibinfo {year} {2024})}\BibitemShut {NoStop}%
\bibitem [{\citenamefont {Berthier}(2014)}]{berthier2014}%
  \BibitemOpen
  \bibfield  {author} {\bibinfo {author} {\bibfnamefont {L.}~\bibnamefont
  {Berthier}},\ }\href {\doibase
  https://doi.org/10.1103/PhysRevLett.112.220602} {\bibfield  {journal}
  {\bibinfo  {journal} {Phys. Rev. Lett.}\ }\textbf {\bibinfo {volume} {112}},\
  \bibinfo {pages} {220602} (\bibinfo {year} {2014})}\BibitemShut {NoStop}%
\bibitem [{\citenamefont {Flenner}\ \emph {et~al.}(2016)\citenamefont
  {Flenner}, \citenamefont {Szamel},\ and\ \citenamefont
  {Berthier}}]{flenner2016}%
  \BibitemOpen
  \bibfield  {author} {\bibinfo {author} {\bibfnamefont {E.}~\bibnamefont
  {Flenner}}, \bibinfo {author} {\bibfnamefont {G.}~\bibnamefont {Szamel}}, \
  and\ \bibinfo {author} {\bibfnamefont {L.}~\bibnamefont {Berthier}},\ }\href
  {\doibase 10.1039/c6sm01322h} {\bibfield  {journal} {\bibinfo  {journal}
  {Soft Matter}\ }\textbf {\bibinfo {volume} {12}},\ \bibinfo {pages}
  {7136–7149} (\bibinfo {year} {2016})}\BibitemShut {NoStop}%
\bibitem [{\citenamefont {Mandal}\ \emph {et~al.}(2016)\citenamefont {Mandal},
  \citenamefont {Bhuyan}, \citenamefont {Rao},\ and\ \citenamefont
  {Dasgupta}}]{mandal2016active}%
  \BibitemOpen
  \bibfield  {author} {\bibinfo {author} {\bibfnamefont {R.}~\bibnamefont
  {Mandal}}, \bibinfo {author} {\bibfnamefont {P.~J.}\ \bibnamefont {Bhuyan}},
  \bibinfo {author} {\bibfnamefont {M.}~\bibnamefont {Rao}}, \ and\ \bibinfo
  {author} {\bibfnamefont {C.}~\bibnamefont {Dasgupta}},\ }\href {\doibase
  10.1039/C5SM02950C} {\bibfield  {journal} {\bibinfo  {journal} {Soft Matter}\
  }\textbf {\bibinfo {volume} {12}},\ \bibinfo {pages} {6268} (\bibinfo {year}
  {2016})}\BibitemShut {NoStop}%
\bibitem [{\citenamefont {Debets}\ \emph {et~al.}(2021)\citenamefont {Debets},
  \citenamefont {Wit},\ and\ \citenamefont {Janssen}}]{debets2021}%
  \BibitemOpen
  \bibfield  {author} {\bibinfo {author} {\bibfnamefont {V.~E.}\ \bibnamefont
  {Debets}}, \bibinfo {author} {\bibfnamefont {X.~M.~D.}\ \bibnamefont {Wit}},
  \ and\ \bibinfo {author} {\bibfnamefont {L.~M.}\ \bibnamefont {Janssen}},\
  }\href {\doibase 10.1103/PhysRevLett.127.278002} {\bibfield  {journal}
  {\bibinfo  {journal} {Phys. Rev. Lett.}\ }\textbf {\bibinfo {volume} {127}},\
  \bibinfo {pages} {278002} (\bibinfo {year} {2021})}\BibitemShut {NoStop}%
\bibitem [{\citenamefont {Mandal}\ and\ \citenamefont
  {Sollich}(2021)}]{mandal2021study}%
  \BibitemOpen
  \bibfield  {author} {\bibinfo {author} {\bibfnamefont {R.}~\bibnamefont
  {Mandal}}\ and\ \bibinfo {author} {\bibfnamefont {P.}~\bibnamefont
  {Sollich}},\ }\href {\doibase 10.1088/1361-648X/abef9b} {\bibfield  {journal}
  {\bibinfo  {journal} {J. Phys.: Condens. Matter}\ }\textbf {\bibinfo {volume}
  {33}},\ \bibinfo {pages} {184001} (\bibinfo {year} {2021})}\BibitemShut
  {NoStop}%
\bibitem [{\citenamefont {Keta}\ \emph {et~al.}(2023)\citenamefont {Keta},
  \citenamefont {Mandal}, \citenamefont {Sollich}, \citenamefont {Jack},\ and\
  \citenamefont {Berthier}}]{keta2023intermittent}%
  \BibitemOpen
  \bibfield  {author} {\bibinfo {author} {\bibfnamefont {Y.-E.}\ \bibnamefont
  {Keta}}, \bibinfo {author} {\bibfnamefont {R.}~\bibnamefont {Mandal}},
  \bibinfo {author} {\bibfnamefont {P.}~\bibnamefont {Sollich}}, \bibinfo
  {author} {\bibfnamefont {R.~L.}\ \bibnamefont {Jack}}, \ and\ \bibinfo
  {author} {\bibfnamefont {L.}~\bibnamefont {Berthier}},\ }\href {\doibase
  10.1039/D3SM00034F} {\bibfield  {journal} {\bibinfo  {journal} {Soft Matter}\
  }\textbf {\bibinfo {volume} {19}},\ \bibinfo {pages} {3871} (\bibinfo {year}
  {2023})}\BibitemShut {NoStop}%
\bibitem [{\citenamefont {Keta}\ \emph {et~al.}(2022)\citenamefont {Keta},
  \citenamefont {Jack},\ and\ \citenamefont {Berthier}}]{keta2022disordered}%
  \BibitemOpen
  \bibfield  {author} {\bibinfo {author} {\bibfnamefont {Y.-E.}\ \bibnamefont
  {Keta}}, \bibinfo {author} {\bibfnamefont {R.~L.}\ \bibnamefont {Jack}}, \
  and\ \bibinfo {author} {\bibfnamefont {L.}~\bibnamefont {Berthier}},\ }\href
  {\doibase 10.1103/PhysRevLett.129.048002} {\bibfield  {journal} {\bibinfo
  {journal} {Phys. Rev. Lett.}\ }\textbf {\bibinfo {volume} {129}},\ \bibinfo
  {pages} {048002} (\bibinfo {year} {2022})}\BibitemShut {NoStop}%
\bibitem [{\citenamefont {Pareek}\ \emph {et~al.}(2025)\citenamefont {Pareek},
  \citenamefont {Sollich}, \citenamefont {Nandi},\ and\ \citenamefont
  {Berthier}}]{pareek2025}%
  \BibitemOpen
  \bibfield  {author} {\bibinfo {author} {\bibfnamefont {P.}~\bibnamefont
  {Pareek}}, \bibinfo {author} {\bibfnamefont {P.}~\bibnamefont {Sollich}},
  \bibinfo {author} {\bibfnamefont {S.~K.}\ \bibnamefont {Nandi}}, \ and\
  \bibinfo {author} {\bibfnamefont {L.}~\bibnamefont {Berthier}},\ }\href
  {\doibase 10.48550/arXiv.2506.09589} {\bibfield  {journal} {\bibinfo
  {journal} {arXiv}\ ,\ \bibinfo {pages} {2506.09589}} (\bibinfo {year}
  {2025})}\BibitemShut {NoStop}%
\bibitem [{\citenamefont {Mandal}\ and\ \citenamefont
  {Sollich}(2020)}]{mandal2020multiple}%
  \BibitemOpen
  \bibfield  {author} {\bibinfo {author} {\bibfnamefont {R.}~\bibnamefont
  {Mandal}}\ and\ \bibinfo {author} {\bibfnamefont {P.}~\bibnamefont
  {Sollich}},\ }\href {\doibase 10.1103/PhysRevLett.125.218001} {\bibfield
  {journal} {\bibinfo  {journal} {Phys. Rev. Lett.}\ }\textbf {\bibinfo
  {volume} {125}},\ \bibinfo {pages} {218001} (\bibinfo {year}
  {2020})}\BibitemShut {NoStop}%
\bibitem [{\citenamefont {Janzen}\ and\ \citenamefont
  {Janssen}(2022)}]{janzen2022aging}%
  \BibitemOpen
  \bibfield  {author} {\bibinfo {author} {\bibfnamefont {G.}~\bibnamefont
  {Janzen}}\ and\ \bibinfo {author} {\bibfnamefont {L.~M.}\ \bibnamefont
  {Janssen}},\ }\href {\doibase 10.1103/PhysRevResearch.4.L012038} {\bibfield
  {journal} {\bibinfo  {journal} {Phys. Rev. Research}\ }\textbf {\bibinfo
  {volume} {4}},\ \bibinfo {pages} {L012038} (\bibinfo {year}
  {2022})}\BibitemShut {NoStop}%
\bibitem [{\citenamefont {Bonn}\ \emph {et~al.}(2002)\citenamefont {Bonn},
  \citenamefont {Tanase}, \citenamefont {Abou}, \citenamefont {Tanaka},\ and\
  \citenamefont {Meunier}}]{bonn2002laponite}%
  \BibitemOpen
  \bibfield  {author} {\bibinfo {author} {\bibfnamefont {D.}~\bibnamefont
  {Bonn}}, \bibinfo {author} {\bibfnamefont {S.}~\bibnamefont {Tanase}},
  \bibinfo {author} {\bibfnamefont {B.}~\bibnamefont {Abou}}, \bibinfo {author}
  {\bibfnamefont {H.}~\bibnamefont {Tanaka}}, \ and\ \bibinfo {author}
  {\bibfnamefont {J.}~\bibnamefont {Meunier}},\ }\href {\doibase
  10.1103/PhysRevLett.89.015701} {\bibfield  {journal} {\bibinfo  {journal}
  {Phys. Rev. Lett.}\ }\textbf {\bibinfo {volume} {89}},\ \bibinfo {pages}
  {015701} (\bibinfo {year} {2002})}\BibitemShut {NoStop}%
\bibitem [{\citenamefont {Di}\ \emph {et~al.}(2011)\citenamefont {Di},
  \citenamefont {Win}, \citenamefont {McKenna}, \citenamefont {Narita},
  \citenamefont {Lequeux}, \citenamefont {Pullela},\ and\ \citenamefont
  {Cheng}}]{di2011signatures}%
  \BibitemOpen
  \bibfield  {author} {\bibinfo {author} {\bibfnamefont {X.}~\bibnamefont
  {Di}}, \bibinfo {author} {\bibfnamefont {K.}~\bibnamefont {Win}}, \bibinfo
  {author} {\bibfnamefont {G.~B.}\ \bibnamefont {McKenna}}, \bibinfo {author}
  {\bibfnamefont {T.}~\bibnamefont {Narita}}, \bibinfo {author} {\bibfnamefont
  {F.}~\bibnamefont {Lequeux}}, \bibinfo {author} {\bibfnamefont {S.~R.}\
  \bibnamefont {Pullela}}, \ and\ \bibinfo {author} {\bibfnamefont
  {Z.}~\bibnamefont {Cheng}},\ }\href {\doibase 10.1103/PhysRevLett.106.095701}
  {\bibfield  {journal} {\bibinfo  {journal} {Phys. Rev. Lett.}\ }\textbf
  {\bibinfo {volume} {106}},\ \bibinfo {pages} {095701} (\bibinfo {year}
  {2011})}\BibitemShut {NoStop}%
\bibitem [{\citenamefont {Ramos}\ and\ \citenamefont
  {Cipelletti}(2001)}]{ramos2001ultraslow}%
  \BibitemOpen
  \bibfield  {author} {\bibinfo {author} {\bibfnamefont {L.}~\bibnamefont
  {Ramos}}\ and\ \bibinfo {author} {\bibfnamefont {L.}~\bibnamefont
  {Cipelletti}},\ }\href {\doibase 10.1103/PhysRevLett.87.245503} {\bibfield
  {journal} {\bibinfo  {journal} {Phys. Rev. Lett.}\ }\textbf {\bibinfo
  {volume} {87}},\ \bibinfo {pages} {245503} (\bibinfo {year}
  {2001})}\BibitemShut {NoStop}%
\bibitem [{\citenamefont {Riechers}\ \emph {et~al.}(2022)\citenamefont
  {Riechers}, \citenamefont {Roed}, \citenamefont {Mehri}, \citenamefont
  {Ingebrigtsen}, \citenamefont {Hecksher}, \citenamefont {Dyre},\ and\
  \citenamefont {Niss}}]{riechers2022predicting}%
  \BibitemOpen
  \bibfield  {author} {\bibinfo {author} {\bibfnamefont {B.}~\bibnamefont
  {Riechers}}, \bibinfo {author} {\bibfnamefont {L.~A.}\ \bibnamefont {Roed}},
  \bibinfo {author} {\bibfnamefont {S.}~\bibnamefont {Mehri}}, \bibinfo
  {author} {\bibfnamefont {T.~S.}\ \bibnamefont {Ingebrigtsen}}, \bibinfo
  {author} {\bibfnamefont {T.}~\bibnamefont {Hecksher}}, \bibinfo {author}
  {\bibfnamefont {J.~C.}\ \bibnamefont {Dyre}}, \ and\ \bibinfo {author}
  {\bibfnamefont {K.}~\bibnamefont {Niss}},\ }\href {\doibase
  10.1126/sciadv.abl9809} {\bibfield  {journal} {\bibinfo  {journal} {Science
  Advances}\ }\textbf {\bibinfo {volume} {8}},\ \bibinfo {pages} {eabl9809}
  (\bibinfo {year} {2022})}\BibitemShut {NoStop}%
\bibitem [{\citenamefont {Kob}\ and\ \citenamefont
  {Barrat}(1997)}]{kob1997aging}%
  \BibitemOpen
  \bibfield  {author} {\bibinfo {author} {\bibfnamefont {W.}~\bibnamefont
  {Kob}}\ and\ \bibinfo {author} {\bibfnamefont {J.-L.}\ \bibnamefont
  {Barrat}},\ }\href {\doibase 10.1103/PhysRevLett.78.4581} {\bibfield
  {journal} {\bibinfo  {journal} {Phys. Rev. Lett.}\ }\textbf {\bibinfo
  {volume} {78}},\ \bibinfo {pages} {4581} (\bibinfo {year}
  {1997})}\BibitemShut {NoStop}%
\bibitem [{\citenamefont {Simha}\ \emph {et~al.}(1984)\citenamefont {Simha},
  \citenamefont {Curro},\ and\ \citenamefont {Robertson}}]{simha1984molecular}%
  \BibitemOpen
  \bibfield  {author} {\bibinfo {author} {\bibfnamefont {R.}~\bibnamefont
  {Simha}}, \bibinfo {author} {\bibfnamefont {J.~G.}\ \bibnamefont {Curro}}, \
  and\ \bibinfo {author} {\bibfnamefont {R.~E.}\ \bibnamefont {Robertson}},\
  }\href {\doibase 10.1002/pen.760241402} {\bibfield  {journal} {\bibinfo
  {journal} {Polymer Engineering \& Science}\ }\textbf {\bibinfo {volume}
  {24}},\ \bibinfo {pages} {1071} (\bibinfo {year} {1984})}\BibitemShut
  {NoStop}%
\bibitem [{\citenamefont {Peter~G}(2009)}]{peter2009spatiotemporal}%
  \BibitemOpen
  \bibfield  {author} {\bibinfo {author} {\bibfnamefont {W.}~\bibnamefont
  {Peter~G}},\ }\href {\doibase 10.1073/pnas.0812418106} {\bibfield  {journal}
  {\bibinfo  {journal} {Proc. Nat. Acad. Sci. (USA)}\ }\textbf {\bibinfo
  {volume} {106}},\ \bibinfo {pages} {1353} (\bibinfo {year}
  {2009})}\BibitemShut {NoStop}%
\bibitem [{\citenamefont {Lubchenko}\ and\ \citenamefont
  {Wolynes}(2004)}]{lubchenko2004theory}%
  \BibitemOpen
  \bibfield  {author} {\bibinfo {author} {\bibfnamefont {V.}~\bibnamefont
  {Lubchenko}}\ and\ \bibinfo {author} {\bibfnamefont {P.~G.}\ \bibnamefont
  {Wolynes}},\ }\href {\doibase 10.1063/1.1771633} {\bibfield  {journal}
  {\bibinfo  {journal} {J. Chem. Phys.}\ }\textbf {\bibinfo {volume} {121}},\
  \bibinfo {pages} {2852} (\bibinfo {year} {2004})}\BibitemShut {NoStop}%
\bibitem [{\citenamefont {Fodor}\ \emph {et~al.}(2016)\citenamefont {Fodor},
  \citenamefont {Nardini}, \citenamefont {Cates}, \citenamefont {Tailleur},
  \citenamefont {Visco},\ and\ \citenamefont {van Wijland}}]{fodor2016}%
  \BibitemOpen
  \bibfield  {author} {\bibinfo {author} {\bibfnamefont {E.}~\bibnamefont
  {Fodor}}, \bibinfo {author} {\bibfnamefont {C.}~\bibnamefont {Nardini}},
  \bibinfo {author} {\bibfnamefont {M.~E.}\ \bibnamefont {Cates}}, \bibinfo
  {author} {\bibfnamefont {J.}~\bibnamefont {Tailleur}}, \bibinfo {author}
  {\bibfnamefont {P.}~\bibnamefont {Visco}}, \ and\ \bibinfo {author}
  {\bibfnamefont {F.}~\bibnamefont {van Wijland}},\ }\href {\doibase
  10.1103/PhysRevLett.117.038103} {\bibfield  {journal} {\bibinfo  {journal}
  {Phys. Rev. Lett.}\ }\textbf {\bibinfo {volume} {117}},\ \bibinfo {pages}
  {038103} (\bibinfo {year} {2016})}\BibitemShut {NoStop}%
\bibitem [{\citenamefont {Ramakrishnan}\ and\ \citenamefont
  {Yussouff}(1979)}]{ramakrishnan1979first}%
  \BibitemOpen
  \bibfield  {author} {\bibinfo {author} {\bibfnamefont {T.}~\bibnamefont
  {Ramakrishnan}}\ and\ \bibinfo {author} {\bibfnamefont {M.}~\bibnamefont
  {Yussouff}},\ }\href {\doibase 10.1103/PhysRevB.19.2775} {\bibfield
  {journal} {\bibinfo  {journal} {Phys. Rev. B}\ }\textbf {\bibinfo {volume}
  {19}},\ \bibinfo {pages} {2775} (\bibinfo {year} {1979})}\BibitemShut
  {NoStop}%
\bibitem [{\citenamefont {Castellani}\ and\ \citenamefont
  {Cavagna}(2005)}]{castellani2005spin}%
  \BibitemOpen
  \bibfield  {author} {\bibinfo {author} {\bibfnamefont {T.}~\bibnamefont
  {Castellani}}\ and\ \bibinfo {author} {\bibfnamefont {A.}~\bibnamefont
  {Cavagna}},\ }\href {\doibase 10.1088/1742-5468/2005/05/P05012} {\bibfield
  {journal} {\bibinfo  {journal} {Journal of Statistical Mechanics: Theory and
  Experiment}\ }\textbf {\bibinfo {volume} {2005}},\ \bibinfo {pages} {P05012}
  (\bibinfo {year} {2005})}\BibitemShut {NoStop}%
\bibitem [{\citenamefont {Reichman}\ and\ \citenamefont
  {Charbonneau}(2005)}]{reichman2005mode}%
  \BibitemOpen
  \bibfield  {author} {\bibinfo {author} {\bibfnamefont {D.~R.}\ \bibnamefont
  {Reichman}}\ and\ \bibinfo {author} {\bibfnamefont {P.}~\bibnamefont
  {Charbonneau}},\ }\href {\doibase 10.1088/1742-5468/2005/05/P05013}
  {\bibfield  {journal} {\bibinfo  {journal} {J. Stat. Mech.: Theor. Exp.}\
  }\textbf {\bibinfo {volume} {2005}},\ \bibinfo {pages} {P05013} (\bibinfo
  {year} {2005})}\BibitemShut {NoStop}%
\bibitem [{\citenamefont {Kim}\ and\ \citenamefont {Latz}(2001)}]{kimlatz}%
  \BibitemOpen
  \bibfield  {author} {\bibinfo {author} {\bibfnamefont {B.}~\bibnamefont
  {Kim}}\ and\ \bibinfo {author} {\bibfnamefont {A.}~\bibnamefont {Latz}},\
  }\href {\doibase 10.1209/epl/i2001-00202-4} {\bibfield  {journal} {\bibinfo
  {journal} {Europhys. Lett.}\ }\textbf {\bibinfo {volume} {53}},\ \bibinfo
  {pages} {660} (\bibinfo {year} {2001})}\BibitemShut {NoStop}%
\bibitem [{\citenamefont {Bouchaud}\ \emph {et~al.}(1996)\citenamefont
  {Bouchaud}, \citenamefont {Cugliandolo}, \citenamefont {Kurchan},\ and\
  \citenamefont {M{\'{e}}zard}}]{bouchaud1996}%
  \BibitemOpen
  \bibfield  {author} {\bibinfo {author} {\bibfnamefont {J.-P.}\ \bibnamefont
  {Bouchaud}}, \bibinfo {author} {\bibfnamefont {L.}~\bibnamefont
  {Cugliandolo}}, \bibinfo {author} {\bibfnamefont {J.}~\bibnamefont
  {Kurchan}}, \ and\ \bibinfo {author} {\bibfnamefont {M.}~\bibnamefont
  {M{\'{e}}zard}},\ }\href {\doibase 10.1016/0378-4371(95)00423-8} {\bibfield
  {journal} {\bibinfo  {journal} {Physica A}\ }\textbf {\bibinfo {volume}
  {226}},\ \bibinfo {pages} {243} (\bibinfo {year} {1996})}\BibitemShut
  {NoStop}%
\bibitem [{\citenamefont {G{\"o}tze}(2009)}]{gotze2009complex}%
  \BibitemOpen
  \bibfield  {author} {\bibinfo {author} {\bibfnamefont {W.}~\bibnamefont
  {G{\"o}tze}},\ }\href@noop {} {\emph {\bibinfo {title} {Complex dynamics of
  glass-forming liquids: A mode-coupling theory}}},\ Vol.\ \bibinfo {volume}
  {143}\ (\bibinfo  {publisher} {Oxford University Press, USA},\ \bibinfo
  {year} {2009})\BibitemShut {NoStop}%
\bibitem [{\citenamefont {Zaccarelli}\ \emph {et~al.}(2001)\citenamefont
  {Zaccarelli}, \citenamefont {Foffi}, \citenamefont {Sciortino}, \citenamefont
  {Tartaglia},\ and\ \citenamefont {Dawson}}]{zaccarelli2001}%
  \BibitemOpen
  \bibfield  {author} {\bibinfo {author} {\bibfnamefont {E.}~\bibnamefont
  {Zaccarelli}}, \bibinfo {author} {\bibfnamefont {G.}~\bibnamefont {Foffi}},
  \bibinfo {author} {\bibfnamefont {F.}~\bibnamefont {Sciortino}}, \bibinfo
  {author} {\bibfnamefont {P.}~\bibnamefont {Tartaglia}}, \ and\ \bibinfo
  {author} {\bibfnamefont {K.~A.}\ \bibnamefont {Dawson}},\ }\href {\doibase
  10.1209/epl/i2001-00395-x} {\bibfield  {journal} {\bibinfo  {journal}
  {Europhys. Lett.}\ }\textbf {\bibinfo {volume} {55}},\ \bibinfo {pages} {157}
  (\bibinfo {year} {2001})}\BibitemShut {NoStop}%
\bibitem [{\citenamefont {Warren}\ and\ \citenamefont
  {Rottler}(2013)}]{warren2013quench}%
  \BibitemOpen
  \bibfield  {author} {\bibinfo {author} {\bibfnamefont {M.}~\bibnamefont
  {Warren}}\ and\ \bibinfo {author} {\bibfnamefont {J.}~\bibnamefont
  {Rottler}},\ }\href@noop {} {\bibfield  {journal} {\bibinfo  {journal} {Phys.
  Rev. Lett.}\ }\textbf {\bibinfo {volume} {110}},\ \bibinfo {pages} {025501}
  (\bibinfo {year} {2013})}\BibitemShut {NoStop}%
\bibitem [{\citenamefont {Klongvessa}\ \emph
  {et~al.}(2022{\natexlab{b}})\citenamefont {Klongvessa}, \citenamefont
  {Ybert}, \citenamefont {Cottin-Bizonne}, \citenamefont {Kawasaki},\ and\
  \citenamefont {Leocmach}}]{klongvessa2022aging}%
  \BibitemOpen
  \bibfield  {author} {\bibinfo {author} {\bibfnamefont {N.}~\bibnamefont
  {Klongvessa}}, \bibinfo {author} {\bibfnamefont {C.}~\bibnamefont {Ybert}},
  \bibinfo {author} {\bibfnamefont {C.}~\bibnamefont {Cottin-Bizonne}},
  \bibinfo {author} {\bibfnamefont {T.}~\bibnamefont {Kawasaki}}, \ and\
  \bibinfo {author} {\bibfnamefont {M.}~\bibnamefont {Leocmach}},\ }\href@noop
  {} {\bibfield  {journal} {\bibinfo  {journal} {J. Chem. Phys.}\ }\textbf
  {\bibinfo {volume} {156}} (\bibinfo {year} {2022}{\natexlab{b}})}\BibitemShut
  {NoStop}%
\bibitem [{\citenamefont {Mandal}\ \emph {et~al.}(2020)\citenamefont {Mandal},
  \citenamefont {Bhuyan}, \citenamefont {Chaudhuri}, \citenamefont {Dasgupta},\
  and\ \citenamefont {Rao}}]{mandal2020_natcomm}%
  \BibitemOpen
  \bibfield  {author} {\bibinfo {author} {\bibfnamefont {R.}~\bibnamefont
  {Mandal}}, \bibinfo {author} {\bibfnamefont {P.~J.}\ \bibnamefont {Bhuyan}},
  \bibinfo {author} {\bibfnamefont {P.}~\bibnamefont {Chaudhuri}}, \bibinfo
  {author} {\bibfnamefont {C.}~\bibnamefont {Dasgupta}}, \ and\ \bibinfo
  {author} {\bibfnamefont {M.}~\bibnamefont {Rao}},\ }\href {\doibase
  10.1038/s41467-020-16130-x} {\bibfield  {journal} {\bibinfo  {journal} {Nat.
  Commun.}\ }\textbf {\bibinfo {volume} {11}},\ \bibinfo {pages} {2581}
  (\bibinfo {year} {2020})}\BibitemShut {NoStop}%
\bibitem [{\citenamefont {Szamel}\ and\ \citenamefont
  {Flenner}(2024)}]{szamel2024}%
  \BibitemOpen
  \bibfield  {author} {\bibinfo {author} {\bibfnamefont {G.}~\bibnamefont
  {Szamel}}\ and\ \bibinfo {author} {\bibfnamefont {E.}~\bibnamefont
  {Flenner}},\ }\href {\doibase 10.1039/d4sm00338a} {\bibfield  {journal}
  {\bibinfo  {journal} {Soft Matter}\ }\textbf {\bibinfo {volume} {20}},\
  \bibinfo {pages} {5237} (\bibinfo {year} {2024})}\BibitemShut {NoStop}%
\bibitem [{\citenamefont {Mandal}\ \emph {et~al.}(2022)\citenamefont {Mandal},
  \citenamefont {Nandi}, \citenamefont {Dasgupta}, \citenamefont {Sollich},\
  and\ \citenamefont {Gov}}]{mandal2022random}%
  \BibitemOpen
  \bibfield  {author} {\bibinfo {author} {\bibfnamefont {R.}~\bibnamefont
  {Mandal}}, \bibinfo {author} {\bibfnamefont {S.~K.}\ \bibnamefont {Nandi}},
  \bibinfo {author} {\bibfnamefont {C.}~\bibnamefont {Dasgupta}}, \bibinfo
  {author} {\bibfnamefont {P.}~\bibnamefont {Sollich}}, \ and\ \bibinfo
  {author} {\bibfnamefont {N.~S.}\ \bibnamefont {Gov}},\ }\href {\doibase
  10.1088/2399-6528/ac9c47} {\bibfield  {journal} {\bibinfo  {journal} {Journal
  of Physics Communications}\ }\textbf {\bibinfo {volume} {6}},\ \bibinfo
  {pages} {115001} (\bibinfo {year} {2022})}\BibitemShut {NoStop}%
\bibitem [{\citenamefont {Farhadifar}\ \emph {et~al.}(2007)\citenamefont
  {Farhadifar}, \citenamefont {R{\"{o}}per}, \citenamefont {Aigouy},
  \citenamefont {Eaton},\ and\ \citenamefont
  {J{\"{u}}licher}}]{farhadifar2007}%
  \BibitemOpen
  \bibfield  {author} {\bibinfo {author} {\bibfnamefont {R.}~\bibnamefont
  {Farhadifar}}, \bibinfo {author} {\bibfnamefont {J.-C.}\ \bibnamefont
  {R{\"{o}}per}}, \bibinfo {author} {\bibfnamefont {B.}~\bibnamefont {Aigouy}},
  \bibinfo {author} {\bibfnamefont {S.}~\bibnamefont {Eaton}}, \ and\ \bibinfo
  {author} {\bibfnamefont {F.}~\bibnamefont {J{\"{u}}licher}},\ }\href
  {\doibase 10.1016/j.cub.2007.11.049} {\bibfield  {journal} {\bibinfo
  {journal} {Curr. Biol.}\ }\textbf {\bibinfo {volume} {17}},\ \bibinfo {pages}
  {2095} (\bibinfo {year} {2007})}\BibitemShut {NoStop}%
\bibitem [{\citenamefont {Fletcher}\ \emph {et~al.}(2014)\citenamefont
  {Fletcher}, \citenamefont {Osterfield}, \citenamefont {Baker},\ and\
  \citenamefont {Shvartsman}}]{fletcher2014}%
  \BibitemOpen
  \bibfield  {author} {\bibinfo {author} {\bibfnamefont {A.~G.}\ \bibnamefont
  {Fletcher}}, \bibinfo {author} {\bibfnamefont {M.}~\bibnamefont
  {Osterfield}}, \bibinfo {author} {\bibfnamefont {R.~E.}\ \bibnamefont
  {Baker}}, \ and\ \bibinfo {author} {\bibfnamefont {S.~Y.}\ \bibnamefont
  {Shvartsman}},\ }\href {\doibase https://doi.org/10.1016/j.bpj.2013.11.4498}
  {\bibfield  {journal} {\bibinfo  {journal} {Biophys. J.}\ }\textbf {\bibinfo
  {volume} {106}},\ \bibinfo {pages} {2291} (\bibinfo {year}
  {2014})}\BibitemShut {NoStop}%
\bibitem [{\citenamefont {Barton}\ \emph {et~al.}(2017)\citenamefont {Barton},
  \citenamefont {Henkes}, \citenamefont {Weijer},\ and\ \citenamefont
  {Sknepnek}}]{barton2017}%
  \BibitemOpen
  \bibfield  {author} {\bibinfo {author} {\bibfnamefont {D.~L.}\ \bibnamefont
  {Barton}}, \bibinfo {author} {\bibfnamefont {S.}~\bibnamefont {Henkes}},
  \bibinfo {author} {\bibfnamefont {C.~J.}\ \bibnamefont {Weijer}}, \ and\
  \bibinfo {author} {\bibfnamefont {R.}~\bibnamefont {Sknepnek}},\ }\href
  {\doibase 10.1371/journal.pcbi.1005569} {\bibfield  {journal} {\bibinfo
  {journal} {PLOS Computational Biology}\ }\textbf {\bibinfo {volume} {13}},\
  \bibinfo {pages} {1} (\bibinfo {year} {2017})}\BibitemShut {NoStop}%
\bibitem [{\citenamefont {Sadhukhan}\ \emph
  {et~al.}(2024{\natexlab{b}})\citenamefont {Sadhukhan}, \citenamefont {Nandi},
  \citenamefont {Pandey}, \citenamefont {Paoluzzi}, \citenamefont {Gov},
  \citenamefont {Dasgupta},\ and\ \citenamefont {Nandi}}]{sadhukhan2024}%
  \BibitemOpen
  \bibfield  {author} {\bibinfo {author} {\bibfnamefont {S.}~\bibnamefont
  {Sadhukhan}}, \bibinfo {author} {\bibfnamefont {M.}~\bibnamefont {Nandi}},
  \bibinfo {author} {\bibfnamefont {S.}~\bibnamefont {Pandey}}, \bibinfo
  {author} {\bibfnamefont {M.}~\bibnamefont {Paoluzzi}}, \bibinfo {author}
  {\bibfnamefont {N.}~\bibnamefont {Gov}}, \bibinfo {author} {\bibfnamefont
  {C.}~\bibnamefont {Dasgupta}}, \ and\ \bibinfo {author} {\bibfnamefont
  {S.~K.}\ \bibnamefont {Nandi}},\ }\href {\doibase 10.1039/d4sm00352g}
  {\bibfield  {journal} {\bibinfo  {journal} {Soft Matter}\ }\textbf {\bibinfo
  {volume} {20}},\ \bibinfo {pages} {6160} (\bibinfo {year}
  {2024}{\natexlab{b}})}\BibitemShut {NoStop}%
\end{thebibliography}

\begin{thebibliography}{8}%
\makeatletter
\providecommand \@ifxundefined [1]{%
 \@ifx{#1\undefined}
}%
\providecommand \@ifnum [1]{%
 \ifnum #1\expandafter \@firstoftwo
 \else \expandafter \@secondoftwo
 \fi
}%
\providecommand \@ifx [1]{%
 \ifx #1\expandafter \@firstoftwo
 \else \expandafter \@secondoftwo
 \fi
}%
\providecommand \natexlab [1]{#1}%
\providecommand \enquote  [1]{``#1''}%
\providecommand \bibnamefont  [1]{#1}%
\providecommand \bibfnamefont [1]{#1}%
\providecommand \citenamefont [1]{#1}%
\providecommand \href@noop [0]{\@secondoftwo}%
\providecommand \href [0]{\begingroup \@sanitize@url \@href}%
\providecommand \@href[1]{\@@startlink{#1}\@@href}%
\providecommand \@@href[1]{\endgroup#1\@@endlink}%
\providecommand \@sanitize@url [0]{\catcode `\\12\catcode `\$12\catcode
  `\&12\catcode `\#12\catcode `\^12\catcode `\_12\catcode `\%12\relax}%
\providecommand \@@startlink[1]{}%
\providecommand \@@endlink[0]{}%
\providecommand \url  [0]{\begingroup\@sanitize@url \@url }%
\providecommand \@url [1]{\endgroup\@href {#1}{\urlprefix }}%
\providecommand \urlprefix  [0]{URL }%
\providecommand \Eprint [0]{\href }%
\providecommand \doibase [0]{http://dx.doi.org/}%
\providecommand \selectlanguage [0]{\@gobble}%
\providecommand \bibinfo  [0]{\@secondoftwo}%
\providecommand \bibfield  [0]{\@secondoftwo}%
\providecommand \translation [1]{[#1]}%
\providecommand \BibitemOpen [0]{}%
\providecommand \bibitemStop [0]{}%
\providecommand \bibitemNoStop [0]{.\EOS\space}%
\providecommand \EOS [0]{\spacefactor3000\relax}%
\providecommand \BibitemShut  [1]{\csname bibitem#1\endcsname}%
\let\auto@bib@innerbib\@empty
\bibitem [{\citenamefont {Nandi}\ and\ \citenamefont
  {Ramaswamy}(2012)}]{nandi2012glassy}%
  \BibitemOpen
  \bibfield  {author} {\bibinfo {author} {\bibfnamefont {S.~K.}\ \bibnamefont
  {Nandi}}\ and\ \bibinfo {author} {\bibfnamefont {S.}~\bibnamefont
  {Ramaswamy}},\ }\href@noop {} {\bibfield  {journal} {\bibinfo  {journal}
  {Phys. Rev. Lett.}\ }\textbf {\bibinfo {volume} {109}},\ \bibinfo {pages}
  {115702} (\bibinfo {year} {2012})}\BibitemShut {NoStop}%
\bibitem [{\citenamefont {Nandi}\ and\ \citenamefont
  {Ramaswamy}(2016)}]{nandi2016glass}%
  \BibitemOpen
  \bibfield  {author} {\bibinfo {author} {\bibfnamefont {S.~K.}\ \bibnamefont
  {Nandi}}\ and\ \bibinfo {author} {\bibfnamefont {S.}~\bibnamefont
  {Ramaswamy}},\ }\href@noop {} {\bibfield  {journal} {\bibinfo  {journal}
  {Phys. Rev. E}\ }\textbf {\bibinfo {volume} {94}},\ \bibinfo {pages} {012607}
  (\bibinfo {year} {2016})}\BibitemShut {NoStop}%
\bibitem [{\citenamefont {Kim}\ and\ \citenamefont {Latz}(2001)}]{kimlatz}%
  \BibitemOpen
  \bibfield  {author} {\bibinfo {author} {\bibfnamefont {B.}~\bibnamefont
  {Kim}}\ and\ \bibinfo {author} {\bibfnamefont {A.}~\bibnamefont {Latz}},\
  }\href {\doibase 10.1209/epl/i2001-00202-4} {\bibfield  {journal} {\bibinfo
  {journal} {Europhys. Lett.}\ }\textbf {\bibinfo {volume} {53}},\ \bibinfo
  {pages} {660} (\bibinfo {year} {2001})}\BibitemShut {NoStop}%
\bibitem [{\citenamefont {Herzbach}(2000)}]{daniel2000}%
  \BibitemOpen
  \bibfield  {author} {\bibinfo {author} {\bibfnamefont {D.}~\bibnamefont
  {Herzbach}},\ }\emph {\bibinfo {title} {Nicht-gleichgewichts-dynamik in
  gl{\"{a}}sern}},\ \href@noop {} {Master's thesis},\ \bibinfo  {school}
  {Institut f{\"{u}}r Physik, Johannes Gutenberg Universit{\"{a}}t, Mainz}
  (\bibinfo {year} {2000})\BibitemShut {NoStop}%
\bibitem [{\citenamefont {Janzen}\ and\ \citenamefont
  {Janssen}(2022)}]{janzen2022aging}%
  \BibitemOpen
  \bibfield  {author} {\bibinfo {author} {\bibfnamefont {G.}~\bibnamefont
  {Janzen}}\ and\ \bibinfo {author} {\bibfnamefont {L.~M.}\ \bibnamefont
  {Janssen}},\ }\href@noop {} {\bibfield  {journal} {\bibinfo  {journal} {Phys.
  Rev. Res.}\ }\textbf {\bibinfo {volume} {4}},\ \bibinfo {pages} {L012038}
  (\bibinfo {year} {2022})}\BibitemShut {NoStop}%
\bibitem [{\citenamefont {Nandi}\ \emph {et~al.}(2018)\citenamefont {Nandi},
  \citenamefont {Mandal}, \citenamefont {Bhuyan}, \citenamefont {Dasgupta},
  \citenamefont {Rao},\ and\ \citenamefont {Gov}}]{nandi2018random}%
  \BibitemOpen
  \bibfield  {author} {\bibinfo {author} {\bibfnamefont {S.~K.}\ \bibnamefont
  {Nandi}}, \bibinfo {author} {\bibfnamefont {R.}~\bibnamefont {Mandal}},
  \bibinfo {author} {\bibfnamefont {P.~J.}\ \bibnamefont {Bhuyan}}, \bibinfo
  {author} {\bibfnamefont {C.}~\bibnamefont {Dasgupta}}, \bibinfo {author}
  {\bibfnamefont {M.}~\bibnamefont {Rao}}, \ and\ \bibinfo {author}
  {\bibfnamefont {N.~S.}\ \bibnamefont {Gov}},\ }\href {\doibase
  10.1073/pnas.1721324115} {\bibfield  {journal} {\bibinfo  {journal} {Proc.
  Nat. Acad. Sci. (USA)}\ }\textbf {\bibinfo {volume} {115}},\ \bibinfo {pages}
  {7688} (\bibinfo {year} {2018})}\BibitemShut {NoStop}%
\bibitem [{\citenamefont {Nandi}\ and\ \citenamefont
  {Gov}(2017)}]{nandi2017nonequilibrium}%
  \BibitemOpen
  \bibfield  {author} {\bibinfo {author} {\bibfnamefont {S.~K.}\ \bibnamefont
  {Nandi}}\ and\ \bibinfo {author} {\bibfnamefont {N.~S.}\ \bibnamefont
  {Gov}},\ }\href {\doibase 10.1039/C7SM01648D} {\bibfield  {journal} {\bibinfo
   {journal} {Soft matter}\ }\textbf {\bibinfo {volume} {13}},\ \bibinfo
  {pages} {7609} (\bibinfo {year} {2017})}\BibitemShut {NoStop}%
\bibitem [{\citenamefont {Warren}\ and\ \citenamefont
  {Rottler}(2013)}]{warren2013quench}%
  \BibitemOpen
  \bibfield  {author} {\bibinfo {author} {\bibfnamefont {M.}~\bibnamefont
  {Warren}}\ and\ \bibinfo {author} {\bibfnamefont {J.}~\bibnamefont
  {Rottler}},\ }\href@noop {} {\bibfield  {journal} {\bibinfo  {journal} {Phys.
  Rev. Lett.}\ }\textbf {\bibinfo {volume} {110}},\ \bibinfo {pages} {025501}
  (\bibinfo {year} {2013})}\BibitemShut {NoStop}%
\end{thebibliography}

%

\end{document}